# Electric-field-induced modulation of thermal conductivity in poly(vinylidene fluoride)

Shichen Deng [1, #], Jiale Yuan [1, #], Yuli Lin [2, #], Xiaoxiang Yu [1], Dengke Ma [3], Yuwen Huang [1], Rencai Ji [1], Guangzu Zhang [2, *], and Nuo Yang [1, *]

[1] State Key Laboratory of Coal Combustion, and School of Energy and Power Engineering, Huazhong University of Science and Technology (HUST), Wuhan 430074, P. R. China

[2] School of Optical and Electronic Information and Wuhan National Laboratory of Optoelectronics, Huazhong University of Science and Technology Wuhan 430074, P. R. China

[3] NNU-SULI Thermal Energy Research Center (NSTER) & Center for Quantum Transport and Thermal Energy Science (CQTES), School of Physics and Technology, Nanjing Normal University, Nanjing, 210023, P. R. China

[#] S. D., J. Y. and Y. L. contributed equally to this work.

E-mail: G.Z. (zhanggz@hust.edu.cn) and N.Y. (nuo@hust.edu.cn)

# ABSTRACT

Phonon engineering focuses on heat transport modulation on atomic-scale. Different from reported methods, it is shown that electric field can also modulate heat transport in ferroelectric polymers, poly(vinylidene fluoride), by both simulation and measurement. Interestingly, thermal conductivities of poly(vinylidene fluoride) array can be enhanced by a factor of 3.25 along the polarization direction by simulation. The semi-crystalline poly(vinylidene fluoride) film can be also enhanced by a factor of 1.5 which is found by both simulation and measurement. The morphology and phonon property analysis reveal that the enhancement arises from the higher inter-chain lattice order, stronger inter-chain interaction, higher phonon group velocity and suppressed phonon scattering. This study offers a new modulation strategy with quick response and without fillers.

# Introduction

Modulating the thermal conductivity of materials is significant in heat dissipation, thermal insulation, thermoelectric and so on. Phonon engineering [1, 2] has arisen in recent decades, which focuses on the modulation of heat transport by utilizing phonon properties like wave effect [3], boundary scattering [4] and anharmonic scattering [5]. Previous works of phonon engineering mainly focus on modulating thermal conductivity of inorganic materials. However, with extensive use of polymers in flexible electronics, photovoltaic energy conversion and high-power electronics [6-8], modulating thermal conductivity of polymers has become an urgent issue duo to the low value of ~ 0.1 $Wm^{-1}K^{-1}$. Moreover, thermal conductive polymers can replace traditional heat spreaders and heat exchanger, with superb features because of good compactness, light-weight and low-cost. Thus, how to modulate the thermal conductivity of polymers is becoming a new focus.

The main strategies of phonon engineering in polymers include engineering morphology in micro/nanoscale and compositing highly thermal conductive fillers [6]. The methods of morphology engineering mainly include mechanical stretching [9, 10], nanoscale-templating [11-13], electrospinning [14, 15] and π-π stacking[16]. Shen et al. [17] and Xu et al. [18] fabricated polyethylene nanofiber and polyethylene film with ultrahigh thermal conductivity, respectively, which validated the MD prediction by Chen [19] on the thermal conductivity of single polymer chain. For polymer nanocomposites, recent efforts mainly focus on reducing interfacial thermal resistance between fillers and polymeric matrix [20], as well as forming inter-filler network [6] to improve the thermal conductivity.

Electric field has been used to modulate physical properties of materials, such as electronic band gap [21], electrical superconductivity [22] and magnetic skyrmions [23]. Modulations with electric field have the advantages of in situ, flexible control of magnitude and direction, quick response and low power consumption [24-26]. The electric-field-induced thermal conductivity modulation has been observed in inorganic materials such as $PbZr_{0.3}Ti_{0.7}O_3$ [27, 28] and silicone [29]. However, electric-field-induced thermal conductivity modulation has rare been observed in polymers. Poly(vinylidene fluoride) (PVDF) is a typical ferroelectric material [30]. It's reported that *β*-phase of PVDF can be poled by electric field [31]. Recently, Dong et al. [32] proposed phonon renormalization induced by electric field in P(VDF-TrFE) chain. However, how can the electric field influence heat transport in in high dimension, which is more typical in applications, need more investigation.

In this work, the electric field effect on thermal conductivity of β-phase PVDF is investigated. First, two kinds of array structure, unpoled array-PVDF (UA-PVDF) and poled array-PVDF (PA-PVDF) are constructed, and their thermal conductivities are calculated with MD simulations. The mechanism of electric field effect on thermal transport is discussed by comparing chain's morphology and phonon properties. Then, P(VDF-TrFE) film is experimentally fabricated and thermal conductivities before and after electric poling are studied, and the results are explained by differential scanning calorimetry (DSC), Fourier transform infrared (FTIR), hysteresis loop and further simulation.

## Methods of simulation, fabrication and measurement

PVDF arrays are constructed by aligning 30 straight PVDF chains, as shown in Figure 1(a), with periodic boundary conditions applied in all three directions. For UA-PVDF, as shown in Figure 1(b), half of chains orientate oppositely with the other half along $y$ direction (direction of dipole moment), the dipole moments of two parts cancel each other out, resulting in zero dipole moment of the whole system. For PA-PVDF, as shown in Figure 1(c), the orientations of all chains along $y$ direction are the same, generating a net dipole moment along y of the whole system. These structures are consistent with previous works [33-35]. Equilibrium molecular dynamics (EMD) simulation method is applied to calculate thermal conductivity [7, 36]. The interactions between atoms are described by the PCFF force field [37]. (Simulation Details are given in S1.) Moreover, thermal conductivity of amorphous PVDF is firstly calculated and compared with experimental values. (Detailed results are given in S2.)

The fabrication of P(VDF-TrFE) films is schematically illustrated in Figure 4(a). P(VDF-TrFE) powders are dissolved into N,N-Dimethylformamide (DMF) under magnetic stirring at room temperature for 24 h to obtain a uniform solution. After stirring, the solution is dropped onto the upper surface of the freshly cleaned glass plate to cast uniformed film with a thickness of 10 μm. The rest of the solvent is driven out by heating at 50 °C for 24 h in the vacuum drying oven and annealed at 75 °C for another 24 h to increase the crystallinity. Then the prepared films are peeled off from the glass plates. For comparing, some films are poled under a DC electric field of 40 and 80 MV/m for 15 min at room temperature. A 10-μm-thick P(VDF-TrFE) film is shown in Figure 4(b).

Thermal conductivities of P(VDF-TrFE) films are measured by 3ω method, and the thermal transport properties of thin films are analyzed by applying bidirectional asymmetric heat transfer model, which has been successfully applied to measure the thermal conductivity of polymer films [38], nanowire array[39] and metal inverse opals [40]. (Details of measurement are given in the S12.)

**Results and discussion**

First, thermal conductivities of UA-PVDF and PA-PVDF at room temperature are evaluated. Before studying the thermal conductivity, the simulation cell size is checked to overcome the finite size effect. (Details are given in S3.) The integrals of HCACF along $x$, $y$, and $z$ directions for two structures at 300K are shown in Figure S3, which demonstrate the anisotropic thermal transport. For UA-PVDF, integral of HCACF converges to 0.16, 0.16, and 24.2 $Wm^{-1}K^{-1}$ along $x$ (inter-chain direction perpendicular to the polarization direction), $y$ (inter-chain direction parallel to the polarization direction) and $z$ (along-chain) directions, respectively. For PA-PVDF, integral of HCACF converges to 0.36, 0.52, and 31.5 $Wm^{-1}K^{-1}$ along $x$, $y$ and $z$ directions, respectively. The anisotropic thermal transport is due to the strong covalent bonds along the chain but weak non-bonded interactions along the inter-chain directions. Compared with UA-PVDF, $\kappa_x$, $\kappa_y$, and $\kappa_z$ of PA-PVDF are increased by 125%, 225% and 30%, respectively. This result indicates that by inducing polarity to the structure, the inter-chain thermal transport (especially along the dipole direction) can be largely enhanced. This improvement is attributed to the changes in structure morphology and phonon properties as will be discussed later.

Figure 1(d) – (f) shows the temperature dependence of thermal conductivities of two structures along three dimensions. The investigated temperature range is 300 K to 500 K. It is noted that the melting point of amorphous PVDF measured experimentally is around 450 K. However, the melting points of polymers increase with chain length and lamellar crystal thickness [41-43]. In simulations of this work, the studied structures of PVDF are crystalline-like, and periodic boundary conditions are applied, which simulates an infinite chain length and lamellar crystal thickness. Therefore, the melting point in systems are higher than 450 K, and the system does not melt at 500 K. Similar phenomena are also observed in MD simulations of other crystalline polymers, such as PE [44], PEO [45] and PEDOT [46]. For UA-PVDF, $\kappa_z$ shows a decreasing trend, but $\kappa_x$ and $\kappa_y$ are temperature independent in the range from 300 K to 500 K. However, for PA-PVDF, thermal conductivities along all directions reduce with temperature. The typical temperature dependence of thermal conductivity in perfect crystal as $T^{-1}$ induced by Umklapp phonon-phonon scattering is plotted as a reference. As can be seen, the decreasing trend of thermal conductivity in PA-PVDF is slightly sharper than the reference line. This is because the morphology changes in polymers are severer than that in perfect crystal at higher temperature[7], which would increase phonon anharmonicity and suppress thermal conductivity. Moreover, the enhancement in thermal conductivity induced by structure poling becomes lower at higher temperature, and almost disappears at 500K.

To explain the thermal conductivity enhancement induced by electric polarization, the structure morphology is investigated. Stable supercell structures after relaxation at 300K for the two structures are shown in Figure 2(a) and 2(b), respectively. Along $z$ direction, all molecular chains are still arranged in a form of array, no obvious along-chain morphological difference is observed between the two structures. The strong covalent bonds dominate the along-chain heat transport, generating high $\kappa_z$ of both structures. And this good alignment also gives $\kappa_z$ the negative temperature dependence which is typical in crystal materials. However, there is a significant discrepancy in inter-chain morphology between UA-PVDF and PA-PVDF. For UA-PVDF, the orientations of molecular chains along $x$ and $y$ directions are inconsistent. Each molecular chain rotates at different angles around its own carbon skeleton, resulting in a different dipole direction of each chain. The inter-chain arrangement of the relaxed structure is more disordered than the initial structure, which can be regarded as an inter-chain amorphous state and would induce numerous phonon scattering [44], resulting in low $\kappa_x$ and $\kappa_y$ of unpoled structure, as well as the temperature-independent inter-chain thermal conductivity. On the contrary, in the relaxed PA-PVDF structure, the orientations of all molecular chains along $x$ and $y$ directions are highly consistent. The inter-chain structure is highly ordered and in an inter-chain crystalline state, which can largely reduce phonon scattering and result in an improved $\kappa_x$ and $\kappa_y$ of PA-PVDF.

To better show the morphology change, inter-chain radial distribution function (RDF) of carbon atoms is calculated for UA-PVDF and PA-PVDF, as is shown in Figure 2(c). (Details of calculating RDF are given in S5.) There are significant changes in RDF of PA-PVDF compared with that of UA-PVDF. First, there are two extra peaks around 8 Å and 12.5 Å, indicating the increase of inter-chain lattice order. Second, peaks shift to the left, suggesting the decrease of inter-chain distance. As a result, the inter-chain nonbonding interactions which are mainly responsible for the inter-chain heat transfer become stronger [47, 48], resulting in increased $\kappa_x$ and $\kappa_y$ of PA-PVDF. Moreover, the decreased inter-chain distance leaves less space for chain bending and rotation, which would suppress scattering of phonons transporting along the chain and result in slightly higher $\kappa_z$ of PA-PVDF. Third, peaks are sharper, indicating the decreased atomic motion range in the inter-chain directions. The calculated mean square displacement (MSD) of carbon atoms presented in Figure 2(d) also shows the suppressed atomic displacement in the inter-chain directions in PA-PVDF (Details of MSD are given in S5.) When displacement of atoms from their equilibrium positions becomes smaller, the anharmonic phonon scattering is reduced, resulting higher inter-chain thermal conductivities of PA-PVDF. However, when temperature rises to 500 K, as shown in Figure S4(a), discrepancies in RDF between two structures almost disappear, corresponding to the similar thermal conductivities of two structures at 500K. The RDF and MSD of fluorine atoms presented in Figure S5 also shows the same tendency.

To gain further understanding of the polarization effect on thermal conductivity, the phonon spectral energy density (SED) is calculated to analyze the difference in heat conduction between two structures. SED at 300K in the full-frequency range is given in Figure S8. Keeping in view that dominant thermal energy carriers in crystalline polymers are usually low-frequency phonons [46], enlarged SED contour maps in the frequency range below 2.5 THz along $\Gamma - X$ and $\Gamma - Y$, and below 8 THz along $\Gamma - Z$ at 300 K are presented in Figure 3. For SED along $\Gamma - Y$ ($y$ direction), obvious changes between UA-PVDF and PA-PVDF can be observed. As denoted by white dashed ellipses, along the direction from the Brillouin zone edge to the center, the acoustic phonon branches of PA-PVDF exhibit clear spectra, whereas those of UA-PVDF show severe broadening and can't be figured out when frequency is over 0.5 THz. This broadening of peaks corresponds to stronger phonon scattering, and shorter phonon relaxation times [49]. Additionally, curves of these branches of PA-PVDF are steeper than those of UA-PVDF, which can also be observed in vDOS shown in Figure S7, indicating higher phonon group velocities in PA-PVDF. Similar changes are also observed in SED profile along $\Gamma - X$ ($x$ direction).

Additionally, it is noticed that the inter-chain SED profile of PA-PVDF becomes fairly flattened when frequency surpasses 2 THz, indicating that high frequency phonons hardly contribute to the inter-chain thermal transport. This flattened shape can be also observed in the inter-chain phonon dispersion of polyethylene [50], which is caused by the weak vdW interactions between polymer chains. For PA-PVDF, phonon group velocities of branches in white dashed ellipses along $\Gamma - Y$ direction are higher than those along $\Gamma - X$ direction, and there are extra phonon branches along $\Gamma - Y$ direction as donated by white solid ellipses. Thus, $\kappa_y$ are higher than $\kappa_x$ in PA-PVDF.

For SED along Γ - Z ($z$ direction), phonons in UA-PVDF and PA-PVDF have much higher group velocities than those along other two directions, resulting a large value of $\kappa_z$ in both structures. Meanwhile, the dispersion curves along Γ - Z of UA-PVDF is apparently broader than that of PA-PVDF. Because the larger inter-chain distance in UA-PVDF makes segments of chains easier to rotate, which would induce more phonon scattering. The morphology-induced phonon scattering hinders along-chain heat transport, and make a smaller $\kappa_z$ in UA-PVDF.

When temperature rises to 500 K, as shown in Figure S10, along the direction of Γ – X and Γ – Y, phonon branches at 500 K of PA-PVDF become more indistinct compared with that at 300 K, indicating stronger inharmonic phonon scattering. Moreover, vDOS at low frequency presented in Figure S7 shows blue shift, which indicating the decrease of phonon group velocity. Thus, $\kappa_x$ and $\kappa_y$ of PA-PVDF show dramatic decrease. However, for UA-PVDF, SED maps along these two directions shows no obvious discrepancy, resulting in temperature-independent $\kappa_x$ and $\kappa_y$ of UA-PVDF. Along Γ – Z, phonon branches of both PA-PVDF and UA-PVDF become broader at 500 K, resulting decreased $\kappa_z$ of both structures. In addition, at 500 K, the similar width of dispersion curves of two structures indicates a similar degree of phonon scattering, which is different from that at room temperature. At 300 K, Umklapp processes is not fully excited, the morphology-induced scattering makes an effect in heat transport, resulting in a smaller $\kappa_z$ of UA-PVDF. However, at high temperature, Umklapp processes become much severer and dominant, making the scattering saturate and cannot be increased more by morphology-induced scattering, resulting in a similar $\kappa_z$ of two structures.

To further illustrate the feasibility of electric poling, P(VDF-TrFE) film is fabricated and its thermal conductivities are measured for both before and after electric poling. The fabrication of P(VDF-TrFE) films is schematically illustrated in Figure 4(a). The content of β-phase and the percentage of crystallinity are two most important factors of PVDF film.

Firstly, the role of TrFE is to increase the content of β-phase, which possesses the best ferroelectric property among all five phases of PVDF [51]. The content of the β-phase is characterized by Fourier transform infrared (FTIR) (Details are given in S10). Figure 4(c) shows the transmittance for a series of wave numbers obtained by FTIR. It is obtained that the relative contents of β-phase of unpoled P(VDF-TrFE) film is 72.1%, which provides the possibility of switching dipoles under electric field. For poled P(VDF-TrFE) film, the content of β-phase is 72.6%, which is similar to that in unpoled film.

Secondly, the crystallinity of P(VDF-TrFE) film is determined according to the differential scanning calorimetry (DSC) (Details are given in S11). The DSC thermograms of P(VDF-TrFE) films are shown in Figure 4(d), from which the crystallinity is obtained as 51.9%.

The thermal conductivities of P(VDF-TrFE) films are measured by 3ω method [52, 53]. As shown in Figure 5(a), thermal conductivity of unpoled P(VDF-TrFE) film is $0.19 \pm 0.06$ Wm$^{-1}$K$^{-1}$ at 300 K. The film is poled under the electric field of 80 MV/m and 40 MV/m. The thermal conductivities of poled film are measured as $0.29 \pm 0.05$ Wm$^{-1}$K$^{-1}$ and $0.26 \pm 0.02$ Wm$^{-1}$K$^{-1}$ (Figure S16), respectively, which is 53% and 37% higher than the unpoled. That is, the increase of thermal conductivity depends on the applied electric field.

By adjusting applied electric field, the P-E (Polarization-Field) loops in Figure 5(b) indicate that the coercive electric field is 55 MV/m (More data given in S13). That is, when the polarization voltage reaches 55 MV/m, all the dipoles can be switched. Since 40 MV/m is lower than the coercive electric field, the dipoles are not fully poled. Then, the thermal conductivity is also between values of the unpoled and the fully polled.

It is shown that electric field can enhance thermal conductivity by both simulation and measurement. In comparison, as is shown in Table 1, the change amplitudes of thermal conductivity in this work are higher than that in inorganic material PbZr$_{0.3}$Ti$_{0.7}$O$_3$ [27, 28] induced by electric field. Besides, it is comparable to or higher than other strategies proposed recently, like modulating with light [54] and temperature [55].

It is noticed that the measured thermal conductivity of poled film is lower than that by simulation. The reason is that the film fabricated is semi-crystalline which contains amorphous part. Thus, a semi-crystalline structure is also simulated by MD as shown in Figure 5(c), whose crystallinity is about 50% (Details given in S15). For of unpoled structure, the calculated thermal conductivity along polarization direction is $0.19 \pm 0.01$ Wm$^{-1}$K$^{-1}$ (Figure 5(a)). After poled, the thermal conductivity rises to $0.28 \pm 0.03$ Wm$^{-1}$K$^{-1}$, which is in good consistence with the experimental value, 0.29 Wm$^{-1}$K$^{-1}$. This provide guidance that samples with a higher crystallinity and chain alignment can get a larger thermal conductivity.

Table 1. Change amplitude of thermal conductivity induced by different strategies

| Material | Strategy | Value (Wm$^{-1}$K$^{-1}$) | Ratio |
| --- | --- | --- | --- |
| PVDF | Electric-field-induced chain rotation | 0.16 – 0.52 (simulation) | 3.25 |
| P(VDF-TrFE) | Electric-field-induced chain rotation | 0.19 – 0.29 (measurement) | 1.53 |
| PbZr$_{0.3}$Ti$_{0.7}$O$_3$ | Electric-field-induced domain switching | 1.20 – 1.07 (measurement) [28] | 1.11 |
| PbZr$_{0.3}$Ti$_{0.7}$O$_3$ | Electric-field-induced domain switching | 1.42 – 1.61 (measurement) [27] | 1.13 |
| Azopolymer | Light-induced phase change | 0.10 – 0.35 (measurement) [54] | 3.50 |
| CNT/Hexadecane | Temperature-induced phase change | 0.19 – 0.29 (measurement) [56] | 3.00 |
| PNIPAM | Temperature-induced phase change | 0.61 – 0.53 (measurement) [55] | 1.13 |

## Conclusion

In summary, it is shown the thermal conductivity can be modulated by electric filed. The thermal transport in both unpoled and poled PVDF have been studied. In MD simulations of PVDF array, the inter-chain thermal conductivity of unpoled PVDF has a low value about 0.16 Wm$^{-1}$K$^{-1}$, and shows no temperature dependence. After being poled by electric field, thermal conductivities along all three directions are enhanced, and show negative temperature dependence, which is typical in crystalline material. The enhancement ratio along the poling direction is the most significant, from 0.16 to 0.52 Wm$^{-1}$K$^{-1}$ at 300 K. As temperature increasing, the enhancement becomes weaker and almost disappears at 500K.

Experimentally, P(VDF-TrFE) film is fabricated and poled using electric fields as high as 80 MV/m. The values of thermal conductivity of poled films are measured as 0.29 Wm$^{-1}$K$^{-1}$ (poled under 80 MV/m) and 0.26 Wm$^{-1}$K$^{-1}$ (40 MV/m), which is 53% and 37% higher than the unpoled respectively. It is also shown that a higher crystallinity and chain alignment as possible can achieve a higher thermal conductivity.

To understand the mechanism of modulation, RDF, MSD and phonon SED are calculated. It is demonstrated that the enhanced thermal conductivity of poled structure rises from increased inter-chain lattice order, stronger inter-chain interaction, higher phonon group velocity and suppressed phonon scattering. And these effects are much weaker at 500 K, which explain the small discrepancy in thermal conductivity between two structures at higher temperature.

Compared with other modulation methods, such as mechanical stretching, cross-linking or filling, the strategy by using electric field can modulate thermal transport of devices without deformations and do not need to sacrifice mechanical properties, which will be of great convenience in practical applications. In addition, the control of electric field is flexible and quick-response, which can modulate thermal conductivity along different direction by changing the direction of electric field applied. This strategy would have broad application prospects in thermal dissipation, thermoelectric, flexible electronics, and so on. Moreover, because the two states, before/after modulation, can correspond to state 0/1. This strategy may also boost a brand-new area of thermal information devices, such as thermal memory or thermal switch.

## Acknowledgements

This work is financially supported by the National Key Research and Development Project of China No. 2018YFE0127800 (NY), the National Natural Science Foundation of China (No. 51972126 and 51772108) (GZ), the Fundamental Research Funds for the Central Universities, HUST (No. 2019kfyRCPY045, 2019kfyRCPY126 and 2018KFYYXJJ052). The authors thank the discussion of Jun Zhou. The authors thank the National Supercomputing Center in Tianjin (NSCC-TJ) and China Scientific Computing Grid (ScGrid) for providing assistance in computations.

**Competing financial interests**: The authors declare no competing financial interests.

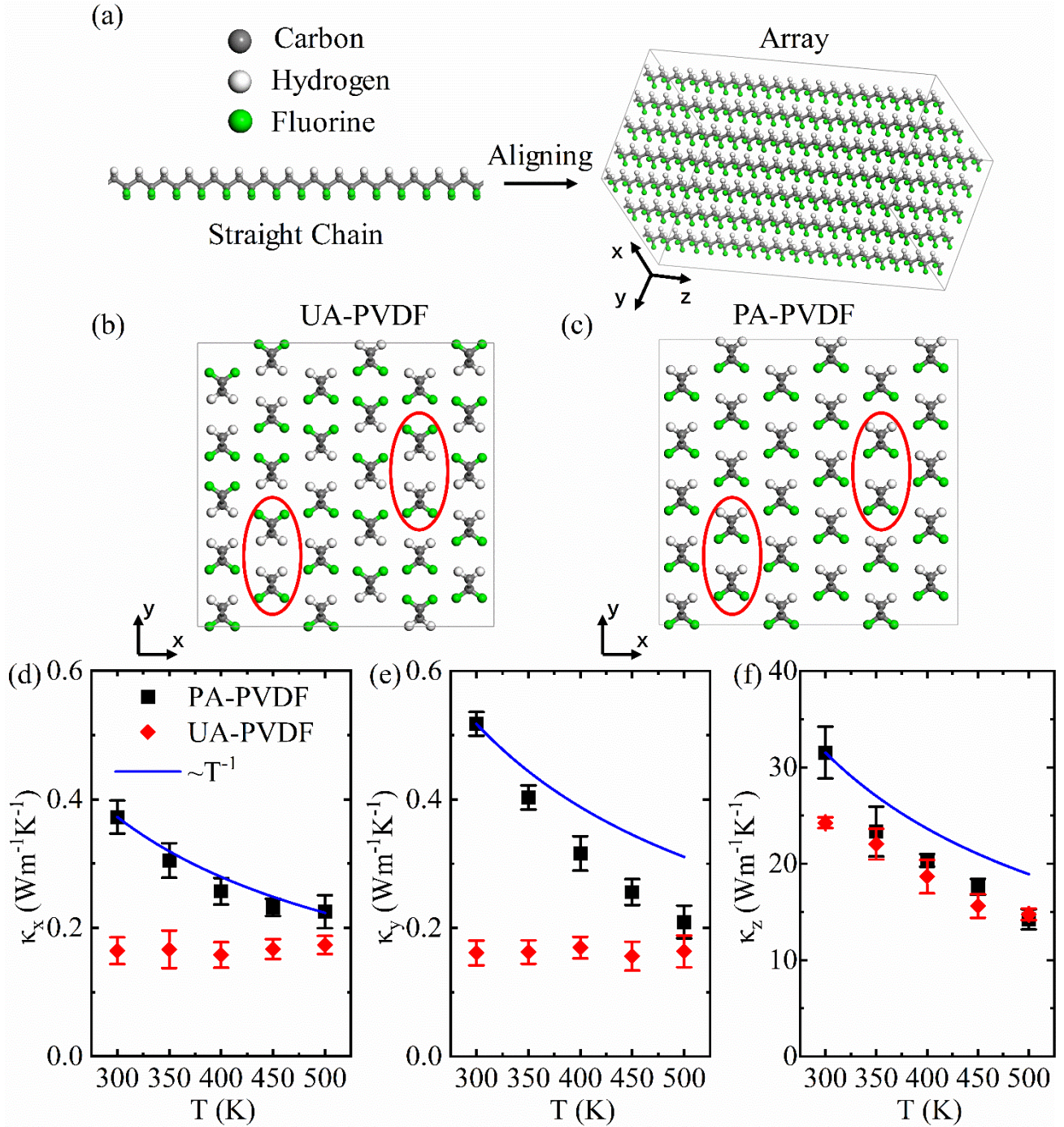

**Figure 1.** Schematics of structure of (a) array-PVDF, (b) unpoled array-PVDF (UA-PVDF), and (c) poled array-PVDF (PA-PVDF). Temperature dependence of thermal conductivity of UA-PVDF and PA-PVDF along directions of (d) x; (e) y and (f) z.

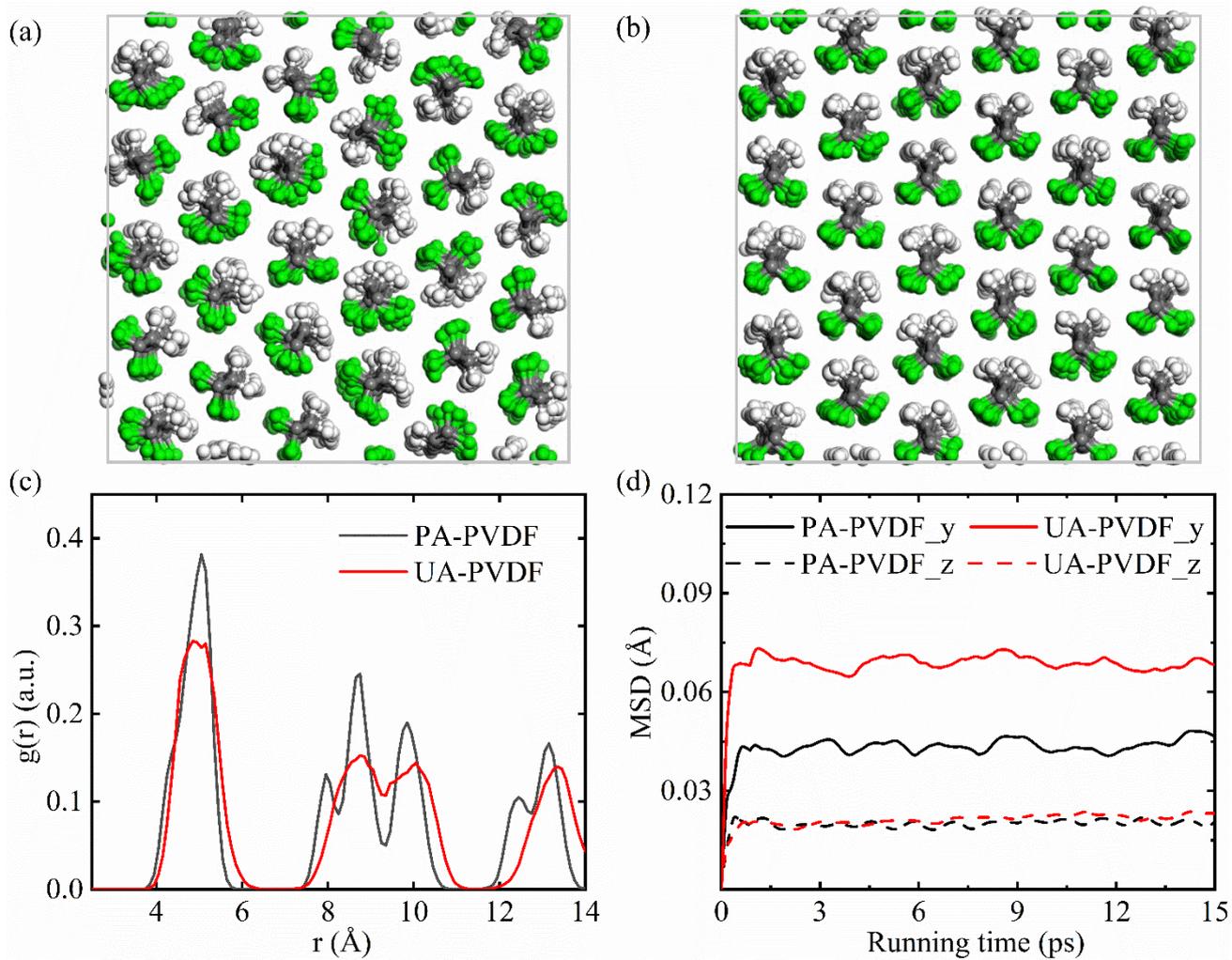

**Figure 2.** Structures after relaxation of (a) UA-PVDF; (b) PA-PVDF at 300 K; (c) Radial distribution function of carbon atoms; (d) mean square displacement of carbon atoms.

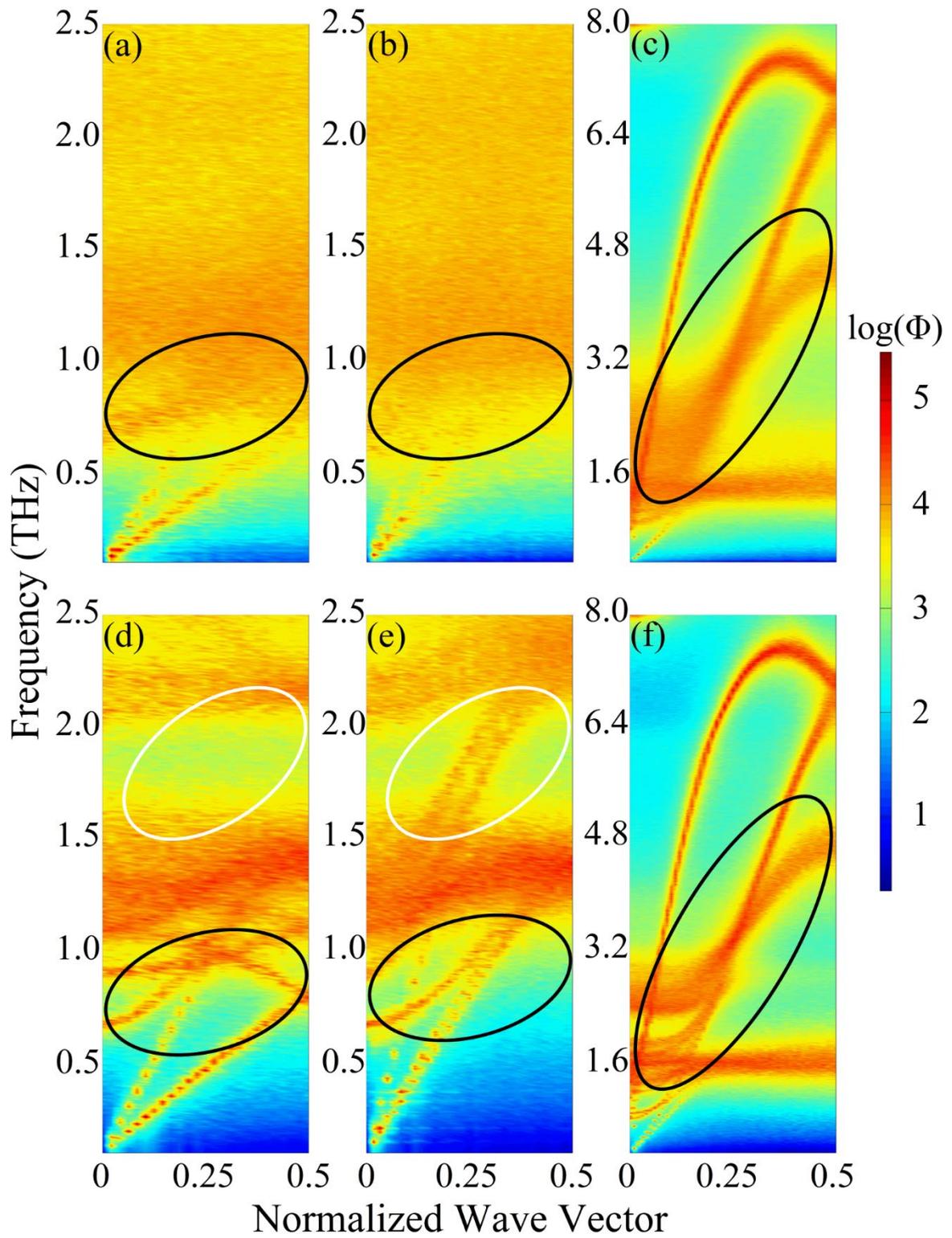

**Figure 3.** SED maps of UA-PVDF along (a) Γ – X, (b) Γ - Y and (c) Γ - Z at 300K; and of PA-PVDF along (d) Γ – X, (e) Γ - Y and (f) Γ - Z below 10 THz at 300K.

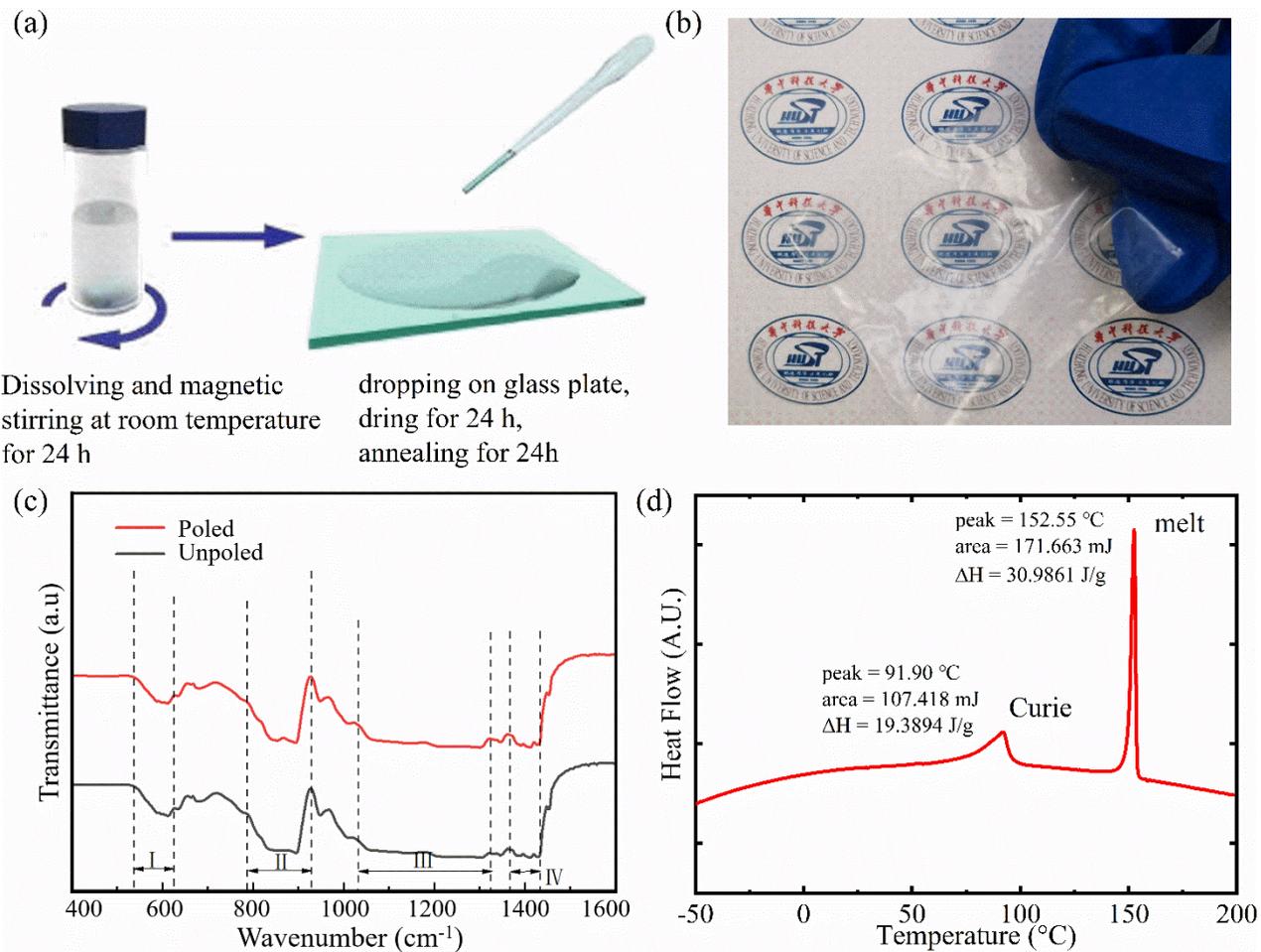

**Figure 4.** (a) Fabrication process of P(VDF-TrFE) films. (b) Photograph of a 10-μm-thick P(VDF-TrFE) film. (c) FTIR spectrums of P(VDF-TrFE) films before and after polarization. (d) The DSC thermogram of P(VDF-TrFE) film.

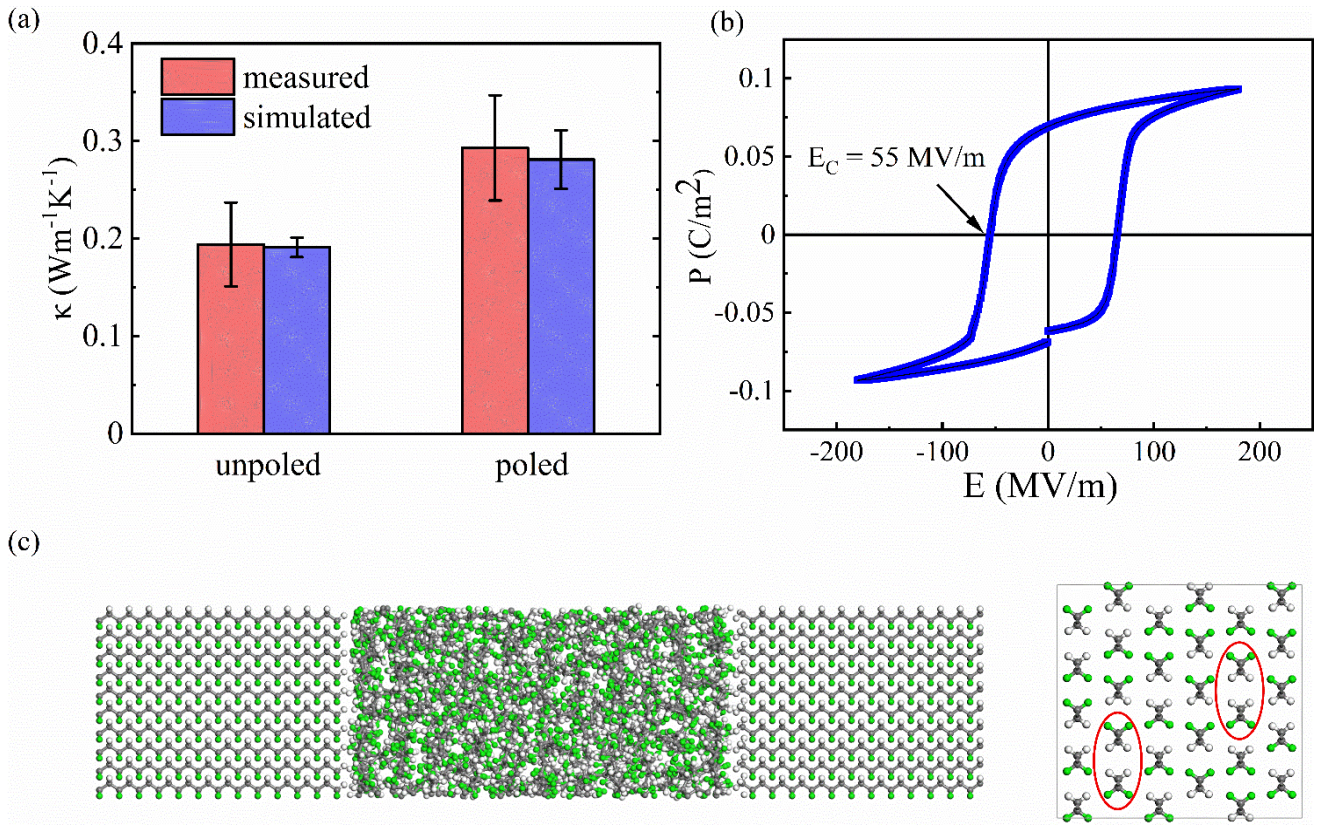

**Figure 5.** (a) Thermal conductivity of unpoled and poled P(VDF-TrFE) films. (b) The PE loop and coercive electric field of P(VDF-TrFE) film. (c) Structure of semi-crystalline PVDF.

**Supporting Information**

# Electric-field-induced thermal conductivity modulation in poly(vinylidene fluoride)

Shichen Deng [1, #], Jiale Yuan [1, #], Yuli Lin [2, #], Xiaoxiang Yu [1], Dengke Ma [3], Yuwen Huang [1], Rencai Ji [1], Guangzu Zhang [2, *], and Nuo Yang [1, *]

[1] State Key Laboratory of Coal Combustion, School of Energy and Power Engineering, Huazhong University of Science and Technology (HUST), Wuhan 430074, P. R. China

[2] School of Optical and Electronic Information and Wuhan National Laboratory of Optoelectronics, Huazhong University of Science and Technology Wuhan 430074, P. R. China

[3] NNU-SULI Thermal Energy Research Center (NSTER) & Center for Quantum Transport and Thermal Energy Science (CQTES), School of Physics and Technology, Nanjing Normal University, Nanjing, 210023, P. R. China

[#] S. D. and X. Y. contributed equally to this work.

Electronic mail: G.Z. (zhanggz@hust.edu.cn) and N.Y. (nuo@hust.edu.cn)

# S1. The EMD simulation details.

The equilibrium molecular dynamics (EMD) simulation method is used to calculate the thermal transport properties. All EMD simulation details are given in Table S1. Green–Kubo formula is a result of the linear response theory and the fluctuation dissipation theorem, which relates the heat flux autocorrelation with the thermal conductivity. Noting that $k_B$ is the Boltzmann constant, V is the system volume, T is the temperature, $\tau$ is the correlation time, $\tau_0$ is the integral upper limit of heat current auto-correlation function (HCACF), E is total kinetic energy of the group of atoms, N is number of total atoms, and the angular bracket denotes an ensemble average. All EMD simulations in this work are performed by the large-scale atomic/molecular massively parallel simulator (LAMMPS) package [1]. The interactions between atoms are described by the polymer consistent force field (PCFF) [43] which includes anharmonic bonding terms and is intended for applications in polymers and organic materials. Scheme S1 lists the partial charges and LJ (12-6) potential parameters of all atom types. The long-range Coulombic force is calculated using the particle−particle particle−mesh (PPPM) method with an error parameter of $10^{-6}$[45, 46]. Periodic boundary conditions are applied in all three dimensions. And the velocity Verlet algorithm is employed to integrate equations of motion [47]. 0.25 fs and 10 Å are chosen as time step and cutoff distance for the Lennard-Jones interaction, respectively. In addition, 5 independent simulations with different initial conditions are conducted to get better average. The simulation structures are simulated in NPT ensembles at target temperatures and 1 atm for 100 ps to obtain the optimized structures and simulation cell sizes, then followed by NVE ensembles for 100 ps before collecting heat flux of all three directions (inter-chain direction x and y, along-chain direction z) in NVE ensembles for 4 ns.

Table S1. Parameter settings in MD simulation.

| Method | EMD (Green-Kubo method) | | | | |
|---|---|---|---|---|---|
| **Force field** | PCFF | | | | |
| **Boundary conditions** | *x y z*: p p p | | | | |
| **Simulation process** | | | | | |
| **Ensemble** | **Settings** | | | | **Purpose** |
| NPT | Time step (fs) | 0.25 | Runtime (ns) | 0.1 | Relax structure |
| | Temperature (K) | 300 | Pressure (atm) | 0 | |

| NVE | Time step (fs) | 0.1 | Runtime (ns) | 0.1 | Relax structure |
|---|---|---|---|---|---|
| | Temperature (K) | 300 | Thermostat | Nose-Hoover | |
| NVE | Sample interval time (fs) | 3 | Runtime (ns) | 4 | Data process |
| | Correlation time (ps) | 100 | Temperature (K) | 300 | |
| **Recorded physical quantity** | | | | | |
| Temperature | $\langle E \rangle = \sum_{i=1}^{N} \frac{1}{2} m_i v_i^2 = \frac{3}{2} N k_B T_{MD}$ | | | | |
| Heat flux | $\vec{J} = \frac{1}{V}\left[\sum_i e_i \vec{v}_i + \frac{1}{2}\sum_i \vec{r}_{ij}(\vec{F}_{ij} \cdot \vec{v}_i)\right]$ | | | | |
| Thermal conductivity | $\kappa = \frac{V}{3 k_B T^2} \int_0^{\tau_0} \langle \vec{J}(0) \cdot \vec{J}(\tau) \rangle d\tau$ | | | | |

-CF2-CH2-
 |    |
 c1   c2

| Atom type | Charge (e) | ε (kcal/mol) | σ (Å) |
|---|---|---|---|
| c2 | -0.106 | 0.054 | 4.01 |
| c1 | 0.5 | 0.054 | 4.01 |
| f | -0.25 | 0.0598 | 3.20 |
| h | 0.053 | 0.02 | 2.995 |

**Scheme S1.** Partial charges and LJ (12-6) potential parameters of all atom types.

## S2. Thermal conductivity of amorphous PVDF.

The structures of amorphous PVDF (A-PVDF) are shown in Figure S1(a). When constructing A-PVDF, a single PVDF chain containing 100 carbon atoms is simulated and equilibrated at 300 K for 1 ns to form a compacted particle. Then 40 of these particles are randomly packed into a supercell. After minimization, a NPT ensemble (a constant number of atoms, pressure and temperature) is used to increase the system temperature from 300 K to 600 K by a constant rate of 50 K/ns, and then a 12 ns NPT run at 600 K is used to generate PVDF melt with fully relaxed amorphous structure. To ensure the amorphous structures obtained are well equilibrated, the average radius of gyration ($R_g$), a quantity to describe the stretch degree of polymer chains, is monitored. As shown in Figure S1(b), $R_g$ gradually increases with the temperature from 300 K to 600 K, which corresponds to the increasing stretch degree of polymer chains. Finally, the oscillating convergence after 11 ns indicates the stable chain morphology, namely the equilibrium amorphous structure. The obtained structure is then quenched to different target temperatures, and NPT ensemble runs for 1 ns are used to further equilibrate the structures at the quenched temperatures. After the stable structures are obtained, the NVE ensemble (constant number of atoms, constant volume, and constant energy) runs for 2 ns are used to record heat flux and calculate thermal conductivity. Due to the isotropic structure of A-PVDF, thermal conductivity along all three directions are used to obtain an average value.

The thermal conductivity of A-PVDF is obtained as $0.22 \pm 0.01$ Wm$^{-1}$K$^{-1}$ at room temperature, which is the same with the reported experimental value of 0.22 Wm$^{-1}$K$^{-1}$ [2]. This value is also similar to that of bulk epoxy [3], cross-linked PE [4] and hydrogel [5]. In addition, it is calculated that thermal conductivity of A-PVDF has no dependence on temperature in the range from 300 K to 600 K, as shown in Figure S1(c). The vibrational density of states (vDOS) is calculated to explain the temperature-independent thermal conductivity, which is shown in Figure S1(d). vDOS includes information about the phonon unharmonic activities. It is found that there is little difference between the vDOS at 300 K, 400 K and 500 K along all three directions. That is, phonon scattering in such amorphous system is already saturated at 300 K and less changed by increasing temperature.

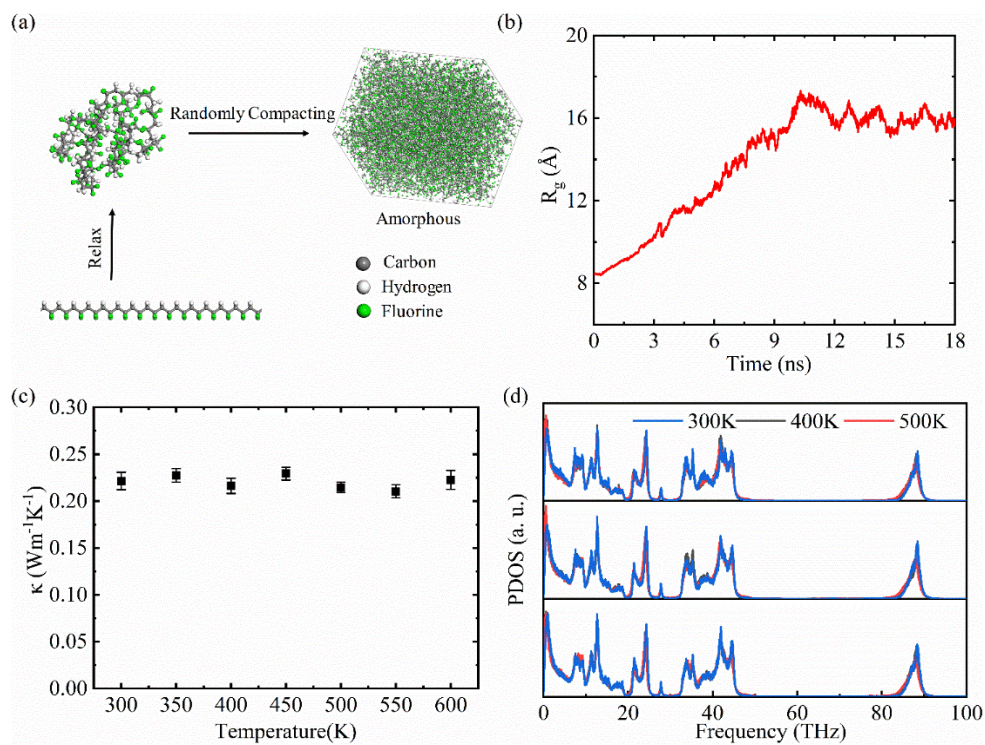

**Figure S1.** (a) Schematics of structure of amorphous PVDF; (b) Average radius of gyration from 40 polymer chains during heat treatment when equilibrate the amorphous PVDF structure; (c) Thermal conductivity of A-PVDF versus temperature; (d) vDOS of A-PVDF along three directions at 300 K, 400K and 500 K.

## S3. The convergence test in size dependence of κ.

When studying the thermal conductivity (κ), the dependence of κ on simulation cell size is checked. When the cell is too small, some phonons, whose wavelengths are longer than the cell size, cannot survived in system. Hence, increasing the domain size will include more long-wavelength phonons in simulation, which contributes to thermal conductivity. When the simulation cell is large enough, the value of thermal conductivity converges due to a competing effect of increasing on both phonon modes and scattering [6].

It is calculated that the dependence of thermal conductivity of PA-PVDF along all directions at 300 K. (1) $L_y$ and $L_z$ are fixed as 1.9 nm and 2.5 nm, and $L_x$ is increased from 1.76 to 4.40 nm. Then, $\kappa_x$ is calculated and show in Figure S2(a). (2) $L_x$ and $L_z$ are fixed as 2.64 nm and 2.5 nm, and $L_y$ is increased from 1.42 to 3.32 nm. Then, $\kappa_y$ is calculated and show in Figure S2(b). (3) $L_x$ and $L_y$ are fixed as 2.37 nm and 1.9 nm, and $L_z$ is increased from 2.5 to 12.5 nm. Then, $\kappa_z$ is calculated and show in Figure S2(c).

As shown in Figure S2, $\kappa_x$, $\kappa_y$ and $\kappa_z$ reach converged values when $L_x$, $L_y$ and $L_z$ are larger than 2.64 nm, 2.37nm and 5nm, respectively. Therefore, to obtain the size independent values of thermal conductivity by EMD, 2.64 nm, 2.37nm and 5nm are chosen for $L_x$, $L_y$ and $L_z$, respectively, in simulations reported in the main text.

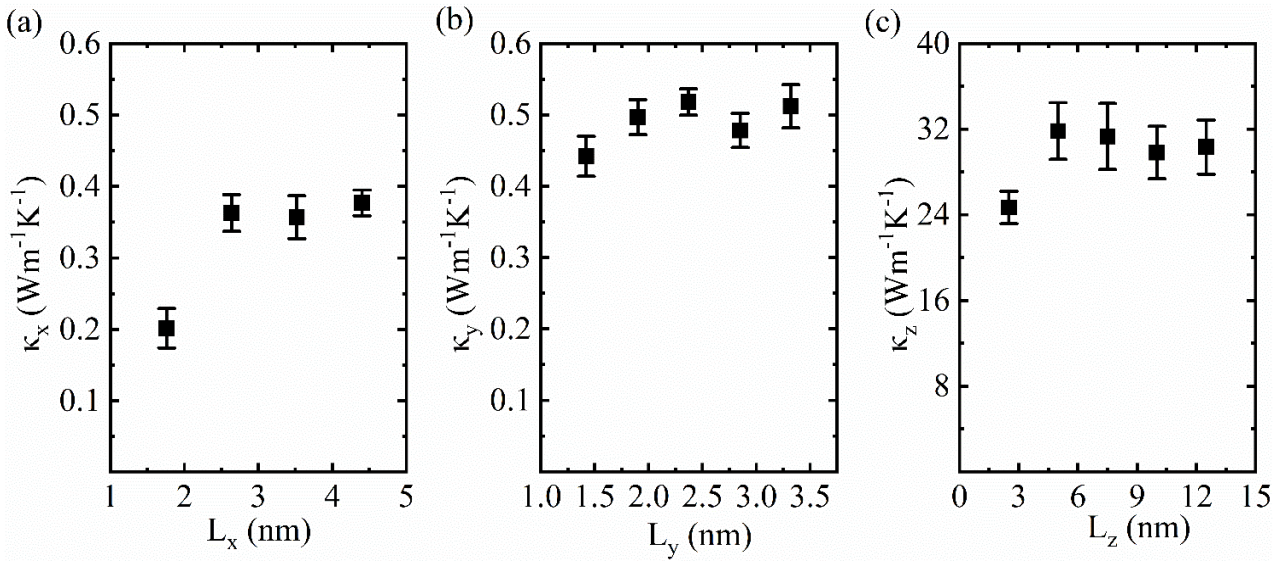

**Figure S2.** Convergence test in size dependence of thermal conductivity of PA-PVDF in (a) *x*; (b) *y* and (c) *z* directions.

## S4. Integral of heat flux auto-correlation function (HCACF) of UA-PVDF and PA-PVDF.

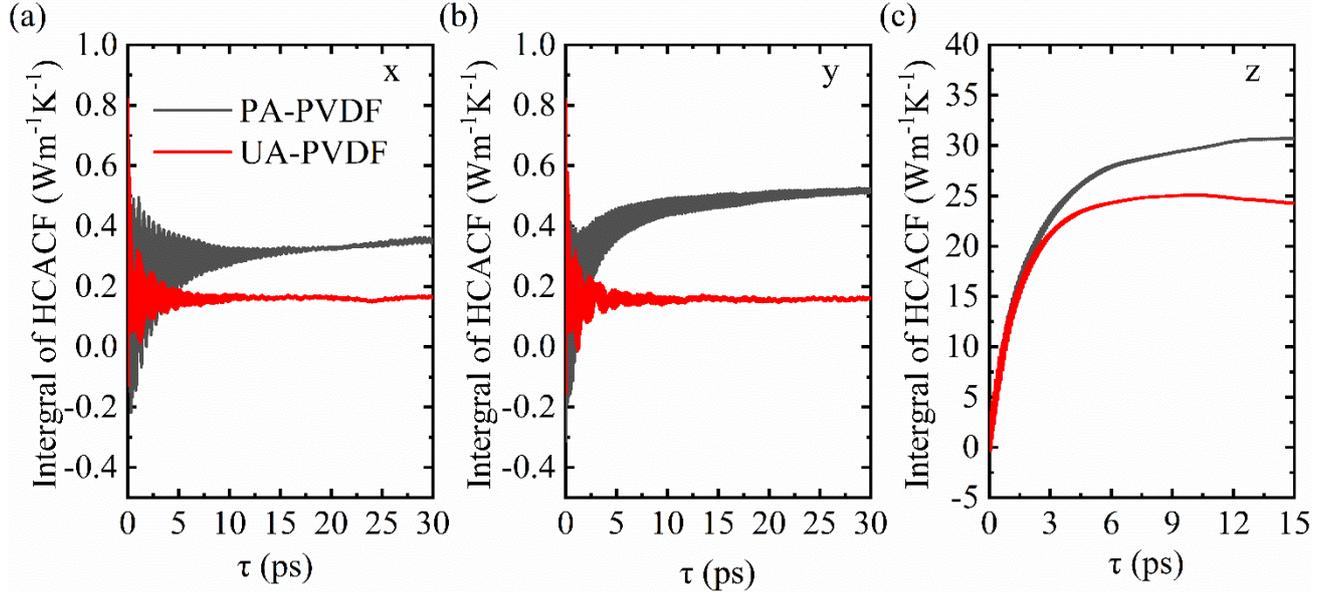

**Figure S3.** Integral of heat flux auto-correlation function (HCACF) in (a) *x*; (b) *y*; (c) z directions of UA-PVDF and PA-PVDF.

## S5. RDF and MSD of C and F atoms.

The inter-chain RDF considers the separations of carbon atoms that are not in the same chain. The reference atom ($x_0$, $y_0$, $z_0$) is the average coordinates of carbon atoms in a single chain. The distance from a reference carbon atom to atoms in other chains is defined as $R = [(x-x_0) + (y-y_0)^2]^{1/2}$, with the pairs satisfying the criterion of $|z-z_0| \leq 2$ Å. This criterion is imposed to reduce noise in the RDF. Then, inter-chain RDF is calculated using $g(r) = n/(2\pi r)$, where n is the number of carbon atoms with a distance of R ($r < R < r + dr$) to the reference atom, and dr is set to be 0.1 Å. Like the three-dimensional RDF, inter-chain RDF reflects the atom density as a function of distance to a reference particle.

MSD of carbon atoms at 300K are shown in Figure S4(b), MSD of carbon atoms along x and y directions of PA-PVDF is obviously reduced compared with that of UA-PVDF. The reduced MSD in PA-PVDF arises from stronger inter-chain interactions which confined the displacement of carbon atoms. The stronger inter-chain interaction origins from the shorter inter-chain distance which has been shown in the inter-chain RDF results (Figure S4(a)). Therefore, the inter-chain thermal transport in PA-PVDF is more efficient. Moreover, in PA-PVDF, MSD along *y* direction is smaller than that along *x* direction, which could explain the larger $\kappa_y$ compared with $\kappa_x$. While in UA-PVDF, discrepancy of MSD along *x* and *y* directions is unobvious, which could explain the similar $\kappa_x$ and $\kappa_y$. Besides, along *z* direction, MSD is the smallest because the strong covalent bonds along chain, and MSD of PA-PVDF shows a slight decrease compared with that of UA-PVDF, which corresponds to the slightly higher along-chain thermal conductivity of PA-PVDF. The MSD of fluorine atoms presented in Figure S5(b) also shows the same tendency.

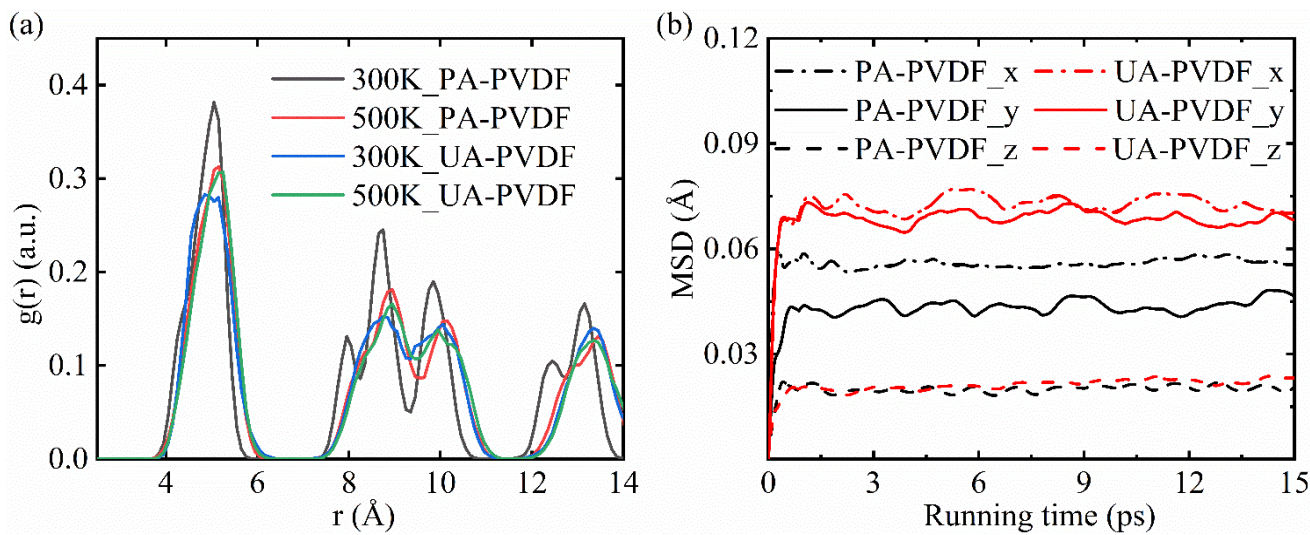

**Figure S4.** (a) RDF of C atoms at 300 K and 500 K; (b) MSD of F atoms at 300K in three directions.

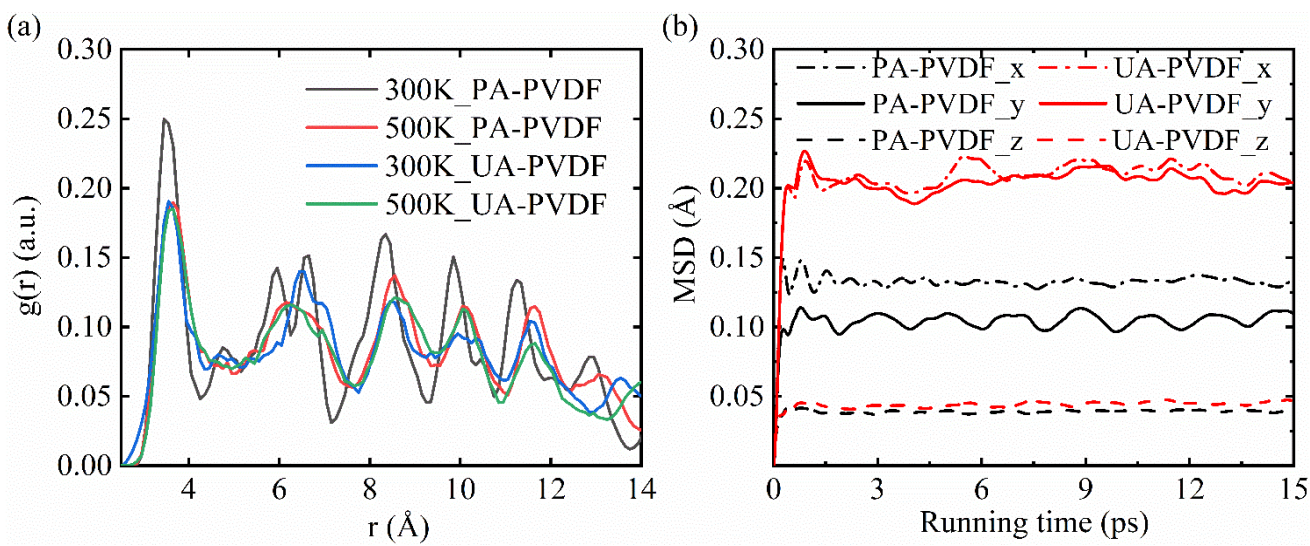

**Figure S5.** (a) RDF of F atoms at 300 K and 500 K; (b) MSD of F atoms at 300K in three directions.

## S6. Phonon density of states for UAPVDF and PAPVDF at 300 K and 500 K.

To find the underlying mechanism of the different thermal conductivities between UAPVDF and PAPVDF, vibrational density of states (vDOS) in three directions at different temperatures is calculated. On the basis of the Parseval's theorem, the mass weighted power spectra $P(\omega)$ are calculated based on velocities. Then vDOS is expressed as $P(\omega)=\frac{1}{N}\sum_{i=1}^{N}m_i|\frac{1}{\sqrt{2\pi}}\int v_i(t)e^{-i\omega t}dt|^2$, here $\omega$ is the angular frequency, $N$ is the number of atoms, $m$ is the atomic mass, $v$ is the velocity, and $t$ is the time interval. We recorded the trajectory of PVDF for 1 nanosecond, then performed fast Fourier transform and multiple time average to obtain the power spectra. vDOS in the full-frequency range is given in Figure S6, and an enlarged vDOS in the frequency range below 10 THz is presented in Figure S7. In $x$ and $y$ directions, low-frequency modes in PAPVDF shift to higher frequency region at 300K. For example, there is a sharp peak at about 1.1 THz in vDOS in $y$ direction of UAPVDF at 300K. When the structure is poled, this peak shows a clear blue-shift to higher frequency about 1.5 THz, which indicates hardening effect and higher group velocity of these phonon modes [7]. Moreover, UAPVDF would induce more phonon scattering. As seen in Figure S7(b), there is an obvious peak at about 2.4 THZ in vDOS in $y$ direction of PAPVDF at 300K. For UAPVDF, however, this peak merges with the peak at lower frequency and almost disappears, which indicates enhanced anharmonic phonon scattering and reduced phonon lifetmes [8]. Based on $\kappa = cv^2\tau/3$, the higher thermal conductivities in $x$ and y directions of PAPVDF is attributed to higher phonon group velocities and longer phonon lifetimes along these two directions. Furthermore, results of vDOS indicate that two effects induced by structure poling mentioned above in y direction are stronger than that in $x$ direction, and the least obvious in $z$ direction. Thus, the enhancement ratio is biggest in y direction and least in $z$ direction.

When temperature rises to 500K, vDOS in all three directions of PA-PVDF shift to low frequency regions, indicating reducing phonon group velocities. And pick width of vDOS is broaden, which means enhanced phonon-phonon scattering induced by increasing temperature. Thus, thermal conductivities of PA-PVDF decrease at higher temperature. Moreover, shape of vDOS of PA-PVDF at 500K is closer with that of UA-PVDF, which can explain the small distinction in thermal conductivities of two structures at high temperature. For UA-PVDF, differences between vDOS at 300K and 500K in $x$ are sufficiently small, resulting similar thermal conductivities at these two temperatures. In y direction, pick around 1.1 THz is sharper at 500 K, but followed by many more flattened picks at higher frequencies. Eventually, these two kinds of changes cancel each other, resulting similar interchain thermal conductivity at different temperatures. Nevertheless, in $z$ direction, most picks at 500K are more flattened than that at 300K in UA-PVDF, indicating more phonon scattering and lower $\kappa_z$ at higher temperature.

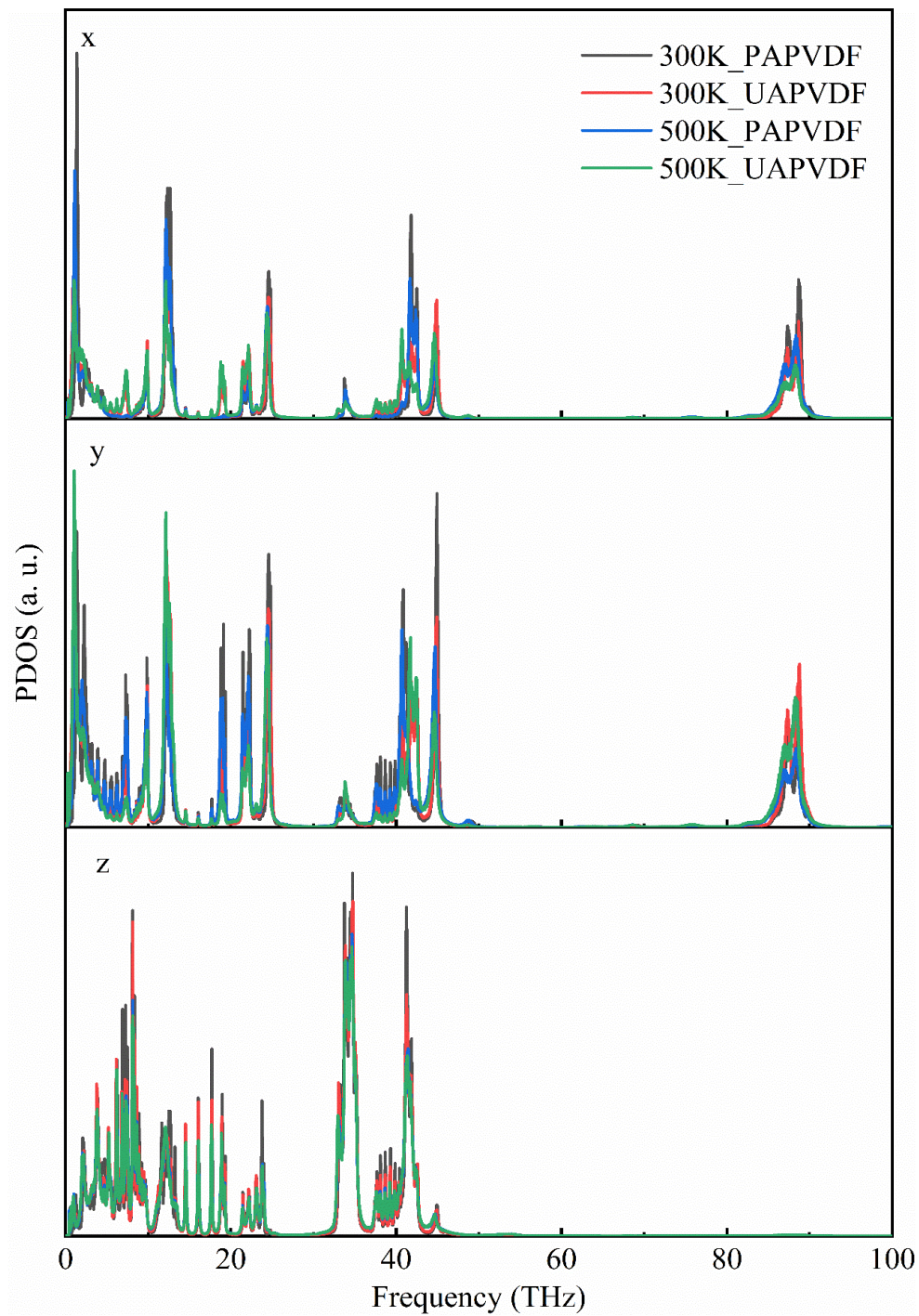

**Figure S6.** Phonon vibrational density of states in *x*, y and *z* direction in full-frequency range.

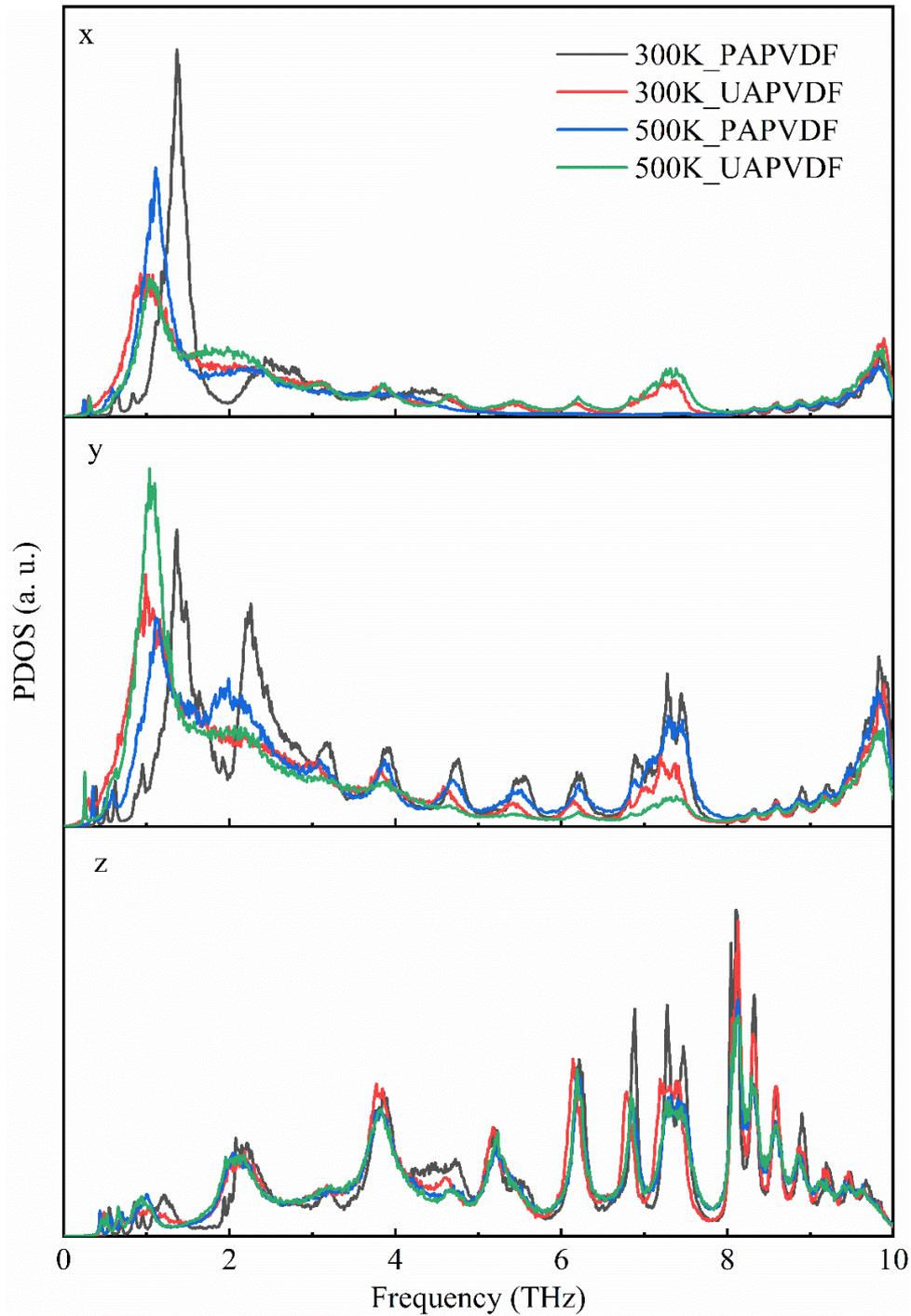

**Figure S7.** Phonon vibrational density of states in *x*, y and *z* direction below 10 THz.

## S7. Spectral energy density in full-frequency range at 300 K.

The SED is evaluated by converting spatial and time-dependent atomic velocity data into frequency and reciprocal space. The SED function Φ(k,ω) is expressed as

$$\Phi(\mathbf{k},\omega) = \frac{1}{4\pi t_{int}N_T}\sum_{a}^{}\sum_{b}^{B}m_b|\int_{0}^{t_{int}}\sum_{n_i}^{N_T}v_a(n_i,b,t)\times \exp[i\mathbf{k}\cdot \mathbf{r}(n_i)-i\omega t]\,dt|^2$$

where $N_T$ is the total number of unit cells, $B$ is the number of atoms per unit cell, and $v_α(n_i,b,t)$ is the velocity in the $α$ direction of atom $b$ (with mass $m_b$) inside of the unit cell $n_i$.

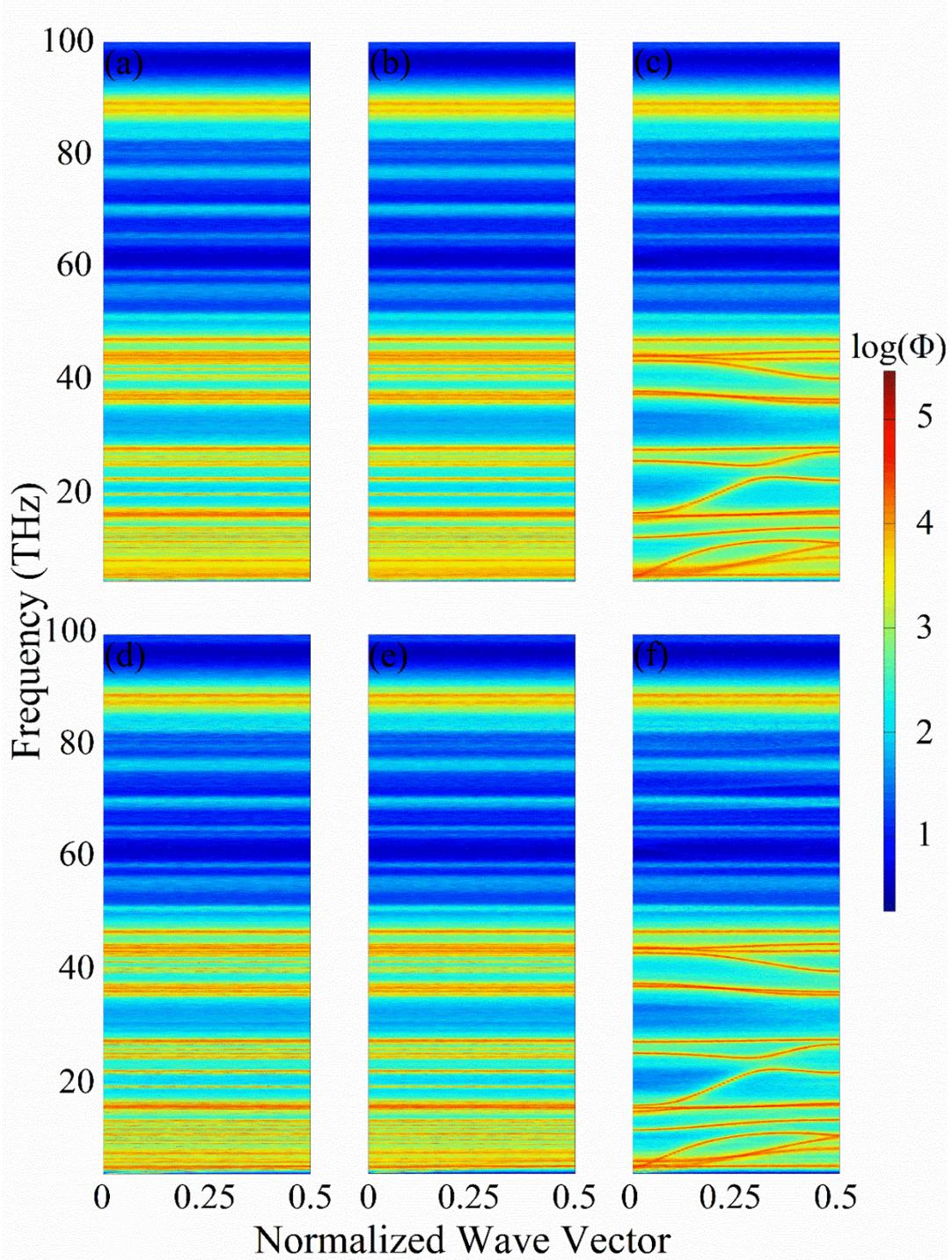

**Figure S8.** SED maps of UA-PVDF along (a) Γ – X, (b) Γ - Y and (c) Γ - Z at 300K; and of PA-PVDF along (d) Γ – X, (e) Γ - Y and (f) Γ - Z in full-frequency range at 300K.

## S8. Spectral energy density at 500 K.

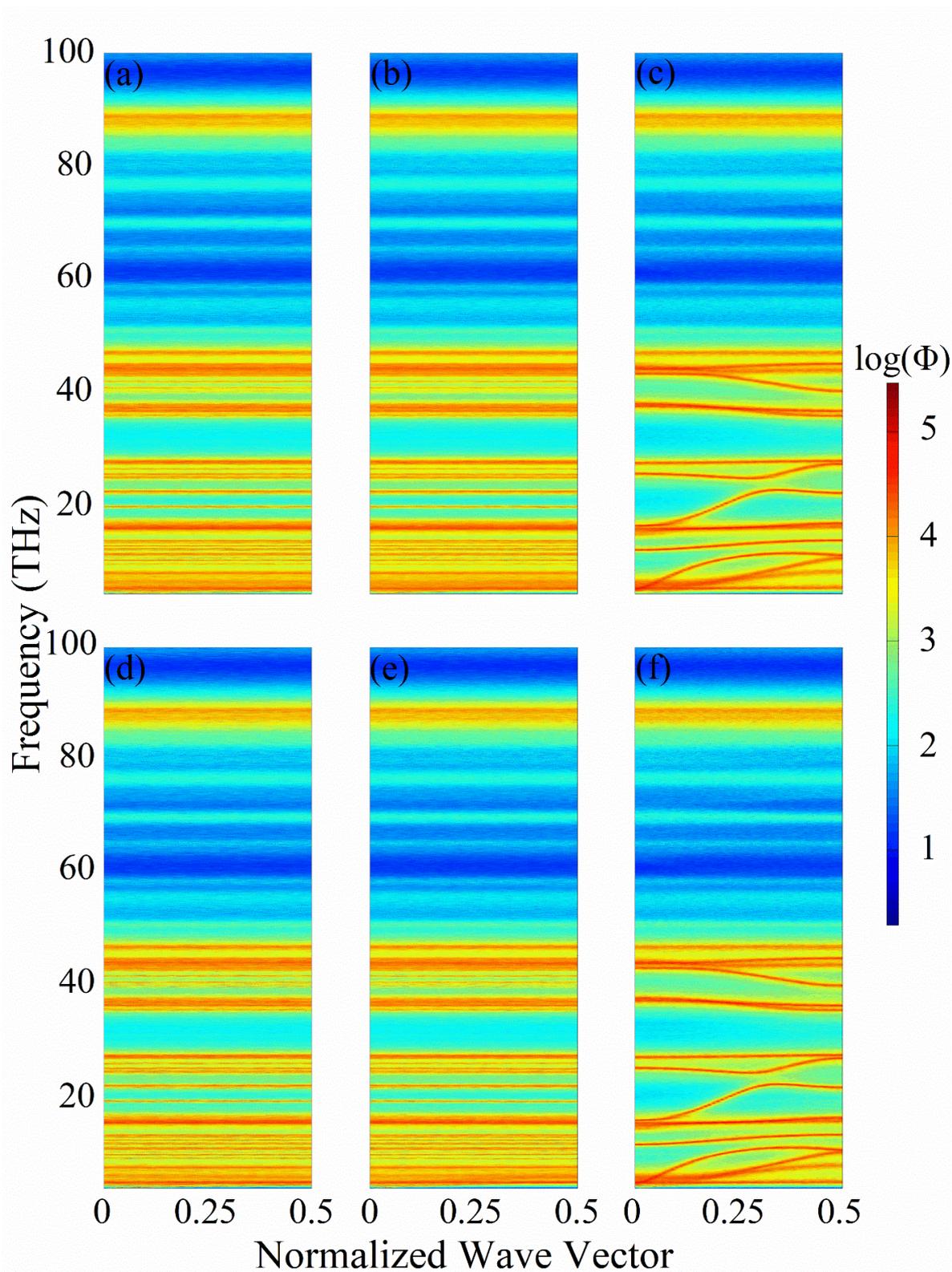

**Figure S9.** SED maps of PA-PVDF along (a) Γ – X, (b) Γ - Y and (c) Γ - Z at 500K; and of UA-PVDF along (d) Γ – X, (e) Γ - Y and (f) Γ - Z in full-frequency range at 500K.

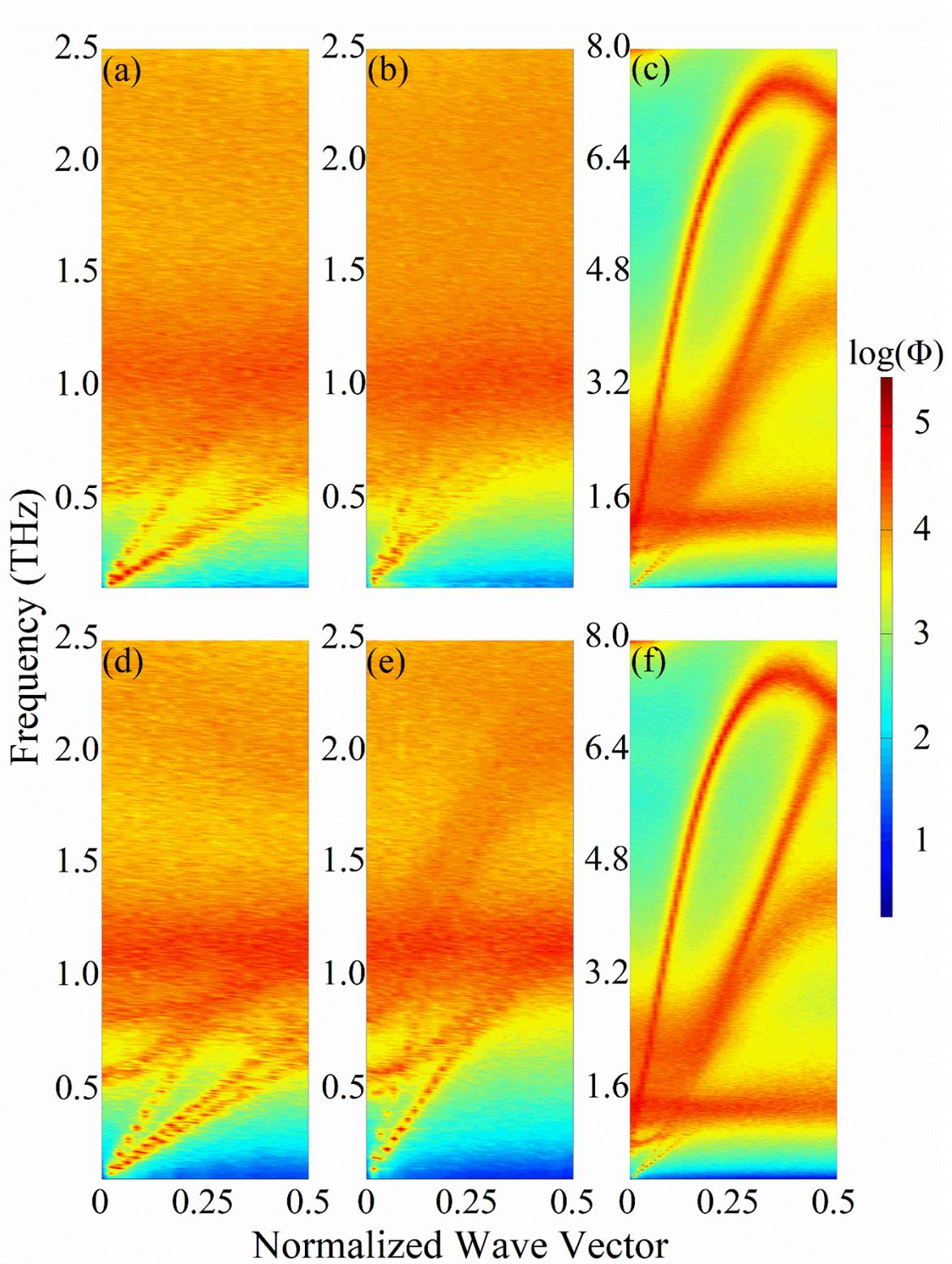

**Figure S10.** SED maps of PA-PVDF along (a) Γ – X, (b) Γ - Y and (c) Γ - Z at 500K; and of UA-PVDF along (d) Γ – X, (e) Γ - Y and (f) Γ - Z below 10 THz at 500K.

## S9. Structure of UA-PVDF after relaxing in different electric fields.

We applied external electric field along *y* direction to the UA-PVDF in MD simulation. With electric field applied, UA-PVDF is simulated in NPT ensembles at 300K and 1 atm for 100 ps, followed by NVE ensembles for 100ps. Then the electric field is removed, followed by the same steps that is used to calculate thermal conductivities. The relaxed structure after simulation in electric field of 1.2 GV/m is shown in Figure S11(a). As a result, the morphology of UA-PVDF is significantly changed by this electric field. All molecular chains and electric dipoles orient well along the direction of electric field, the inter-chain lattice order is largely improved, indicating the polarization of system. The action of molecular chain rotation can be observed in the attached video.

The calculated $\kappa_x$, $\kappa_y$ and $\kappa_z$ after relaxing at different electric field intensities are shown in Figure S11(b), S11(c) and S11(d), respectively. The threshold electric field for polarization is about 0.8 GV/m, which is consistent with the value of other numerical simulations [9, 10] and the experimental results of ultrathin P(VDF-TrFE) films [11, 12]. When electric field intensity is below 0.8 GV/m, inter-chain thermal conductivities is the same with that of initial unpoled structure, and hardly change with electric field. However, the inter-chain thermal conductivities suddenly increase to the value of poled structure when electric field intensity exceeds 0.8 GV/m. To explain the switch behavior of thermal conductivity, it is compared that relaxed structures of UA-PVDF under different electric field intensities, as shown in Figure S12. It is clear that the structure relaxed below 0.8 GV/m has a poor inter-chain arrangement. On the contrary, when applying electric field over 0.8 GV/m, molecular chains in UA-PVDF reorient to the same direction, and the inter-chain lattice order is improved. This switch behavior of morphology leads to similar switch behavior of thermal conductivities. And over the threshold electric field, the morphology would not have noticeable change. Thus, thermal conductivities show no obvious tendency over the threshold electric field.

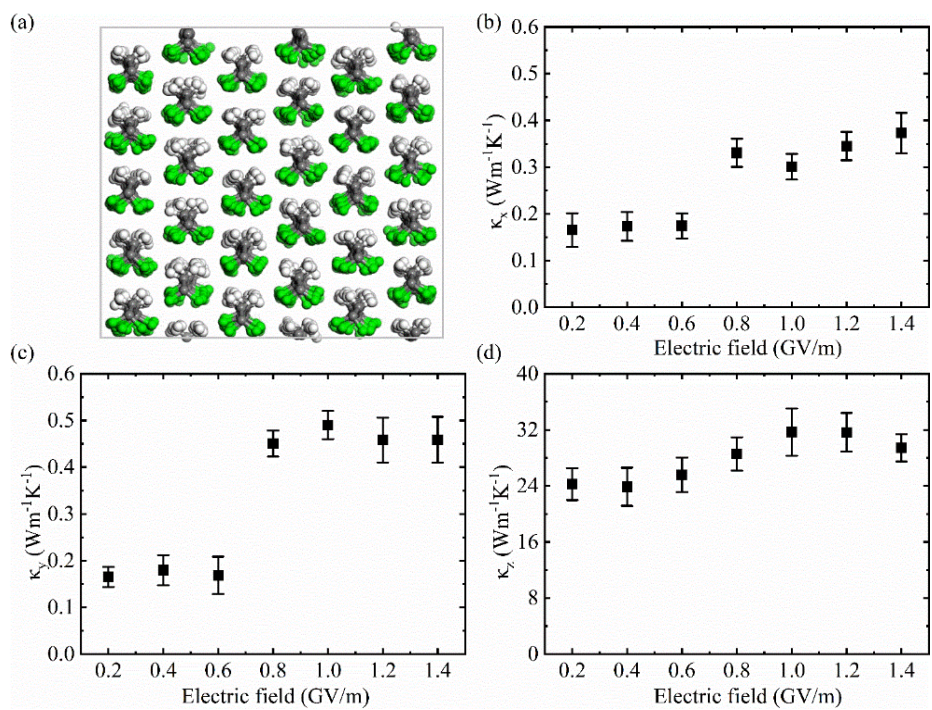

**Figure S11.** (a) Structure of UA-PVDF after relaxing in electric field of 1.2 GV/m; thermal conductivities along (b) *x*; (c) *y*; (d) *z* directions after relaxing in electric field of different intensities.

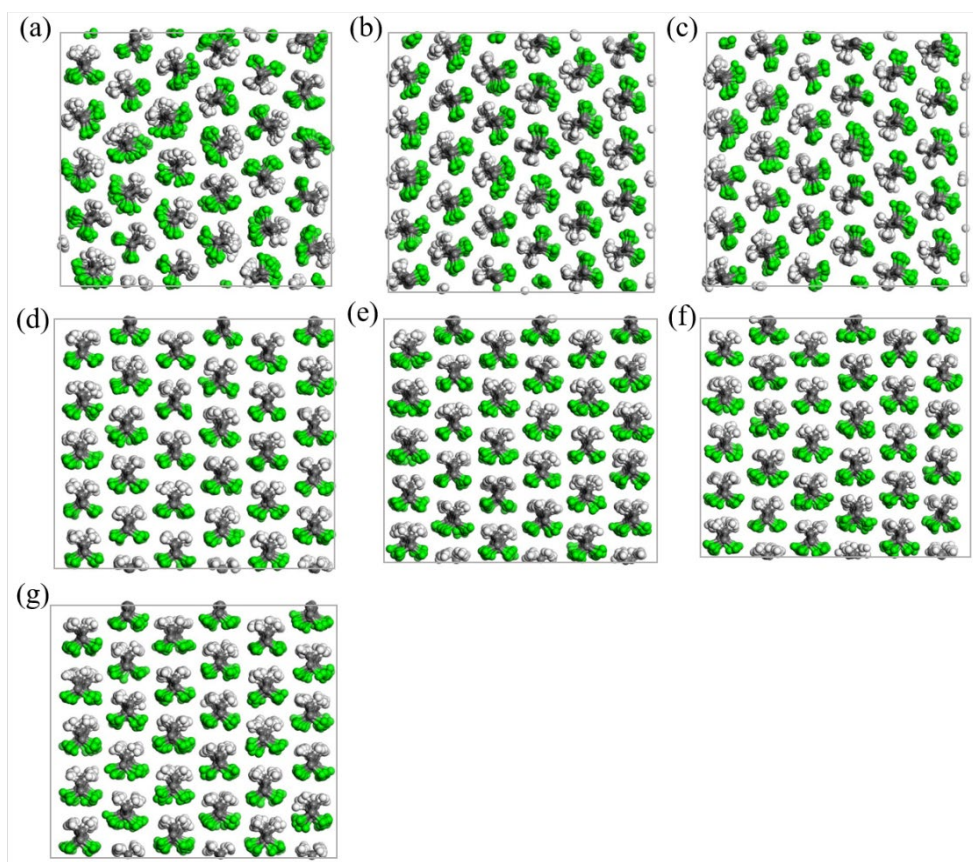

**Figure S12.** Structure of UA-PVDF after relaxing in electric field with intensities of (a) 0.2; (b) 0.4; (c) 0.6; (d) 0.8; (e) 1.0; (f) 1.2; (g) 1.4 GV/m."

## S10. FTIR spectrums of P(VDF-TrFE) films.

PVDF is a kind of ferroelectric polymer, which owns five crystalline structures that are α-, β-, γ-, δ-, and ε-phase[13]. The β-phase PVDF with the dipolar moment of ~8×10$^{-30}$ cm per unit cell has the best ferroelectric property among the five crystalline phases[14] and is most commonly obtained by mechanical stretching of the α-phase[15], from melt under specific conditions[16] or from solution crystallization[17]. Recently, with the aim to improve the content of β-phase, a variety of copolymers of PVDF has been developed[18]. P(VDF-TrFE) is one of the most studied copolymers, which can transform a lot of α-phase to β-phase.

The content of the β-phase in P(VDF-TrFE) films is characterized by Fourier transform infrared (FTIR). Figure 4(c) is the corresponding transmittance for a series of wave numbers. The relative contents of β-phase in P(VDF-TrFE) film can be calculated by the following equation[19].

$$F_{EA} = \frac{A_{EA}}{\left(\frac{K_{841}}{K_{764}}\right) A_{764} + A_{EA}} \times 100\%$$

where $A_{EA}$ and $A_{764}$ are the absorbance intensities at 841 and 764 cm$^{-1}$, respectively, and $K_{841}$ (7.7 × 10$^4$ cm$^2$mol$^{-1}$) and $K_{764}$ (6.1 × 10$^4$ cm$^2$mol$^{-1}$) are the absorption coefficients at the respective wavenumbers. The calculation result shows that 72.1% of β-phase exists in the crystalline structures of unpoled P(VDF-TrFE) film, which provides the possibility of switching domains under electric field. For poled P(VDF-TrFE) film, the content of β-phase is 72.6%, which is similar to that in unpoled film.

## S11. DSC of P(VDF-TrFE) films.

Differential scanning calorimetry (DSC) measurement is performed to determine the crystallinity of P(VDF-TrFE) film. The apparatus was calibrated in both temperature and enthalpy using indium, tin, and mercury standards. DSC trace 70/30 P(VDFTrFE) sample is presented in Figure 4(d). The heating ramp features two endothermic events. The first one situat at 91.90 °C, which is associated with the Curie transition [20]. The second peak occurs at 152.55 °C, corresponds to the melting of it. Degree of crystallinity can be estimated as follows:

$$\chi_C(\%) = \frac{\Delta H_{melt} + \Delta H_{Curie}}{\Delta H_\infty} \times 100$$

where $\Delta H_{melt}$ is the enthalpy of fusion, $\Delta H_{Curie}$ is the enthalpy of the Curie transition, and $\Delta H_\infty$ is the theoretical enthalpy of fusion of 100% crystalline P(VDF-TrFE). $\Delta H_{melt}$ and $\Delta H_{Curie}$ determined from DSC is 30.99 and 19.39 Jg$^{-1}$, respectively. And $\Delta H_\infty$ of 70/30 P(VDF-TrFE) is 97.10 Jg$^{-1}$ [20]. Thus, the crystallinity is obtained as 51.9%.

## S12. 3-Omega method and heat transfer model.

The thermal conductivities of P(VDF-TrFE) films are measured by the 3ω method[21, 22]. The 3ω method can measure a wide range of samples, such as solid bulk[23], liquid[24], thin film[25-27] and one-dimensional materials [28, 29]. As an electrothermal frequency-domain measurement, 3ω method simultaneously generate Joule heating and detect the electrical signals caused by temperature rise. Figure S13(a) shows the diagram of the 3ω measurement platform. In brief, the heater is driven by an alternating current at frequency ω. The Joule heating causes the temperature oscillation at frequency 2ω, and so does the resistance of the heater. The oscillating current and resistance generate a voltage signal at frequency 3ω, which can be detected by measurement platform. In this work, The thermal transport properties of P(VDF-TrFE) films were analyzed by bidirectional asymmetric heat transfer model, which has been successfully used to measure the thermal conductivity of polymer films [30], nanowire array [31] and metal inverse opals [27]. Figure S13(b) is the cross-section view of the heater covered with P(VDF-TrFE) film (Figure S13(d)), which shows the diagram of bidirectional asymmetric heat transfer model.

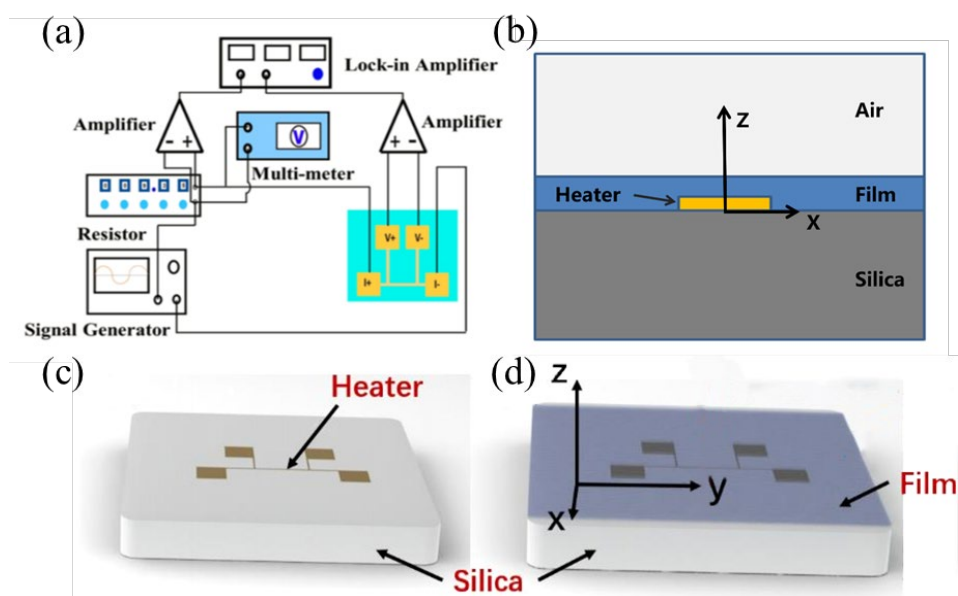

**Figure S13.** The schematic diagram of (a) the 3ω measurement platform, (b) the heat transfer model, (c) the heater, (d) the heater covered with P(VDF-TrFE) film

In the 3-Omega measurement, we obtained the electrode temperature rise in the frequency range of 5-500 Hz. Since the unpoled P(VDF-TrFE) film has a lower thermal conductivity, the temperature rise is higher than the poled.

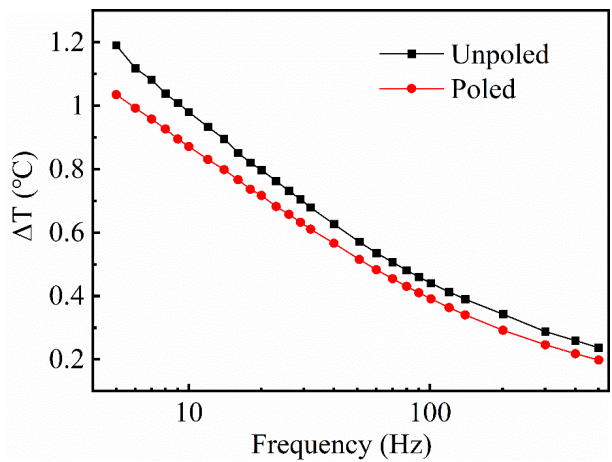

**Figure S14.** Temperature rise of unpoled and poled P(VDF-TrFE) films

## S13. PE loops of P(VDF-TrFE) films under a series of electric field.

Due to the existence of the ferroelectric domain structure, when a strong alternating electric field is applied to the ferroelectric material, the hysteresis effect of the electric domain will generate a ferroelectric hysteresis loop. The P-E (Polarization-Field) loops are an external manifestation of the existence of ferroelectric domain structures in materials. PE loops in Figure S15 indicate that the coercive electric field is 55 MV/m. That is, when the polarization voltage reaches 55 MV/m, all dipoles can be switched.

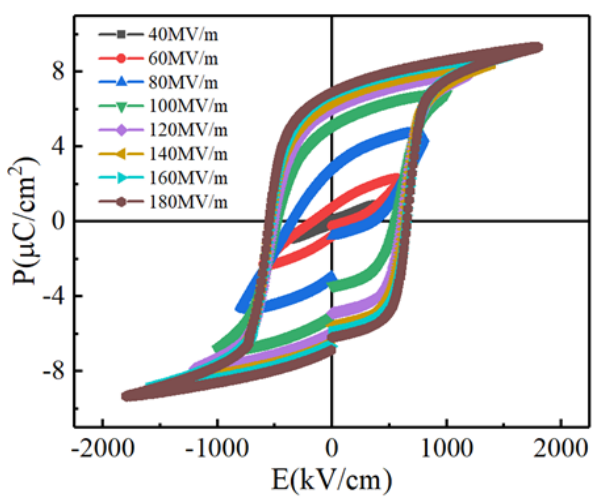

**Figure S15.** PE loops of P(VDF-TrFE) films under a series of electric field.

# S14. Thermal conductivity of P(VDF-TrFE) films after poled in different electric fields.

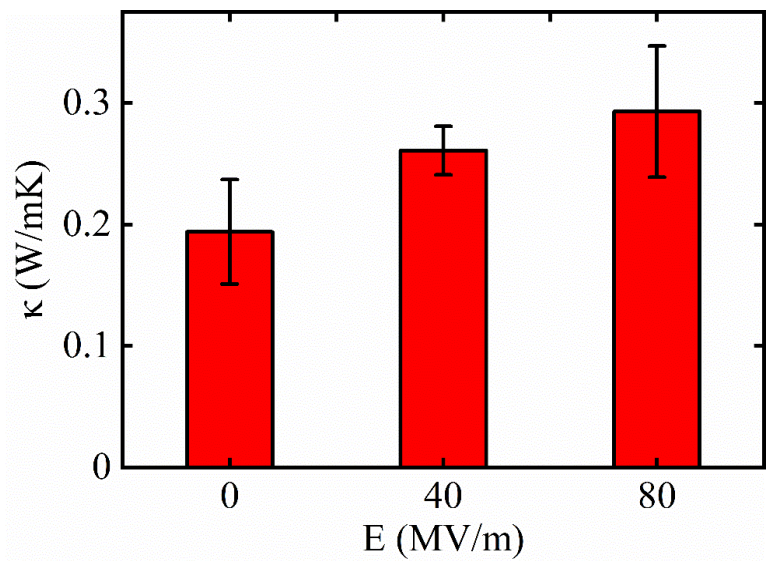

**Figure S16.** Thermal conductivity of P(VDF-TrFE) films after poled in different electric fields.

# S15. MD simulation details of semi-crystalline PVDF structure.

The initial structures of semi-crystalline PVDF are shown in Figure S17(a) and S17(b). When constructing the amorphous part, a single PVDF chain containing 50 carbon atoms is simulated and equilibrated at 300 K for 1 ns to form a compacted particle. Then 6 particles are randomly packed into a supercell. After minimization, an NPT ensemble is used to increase the system temperature from 300 K to 600 K by a constant rate of 75 K/ns, and then a 9 ns NPT run at 600 K is used to generate PVDF melt with fully relaxed amorphous structure. The obtained structure is then quenched to 300K, and NPT ensemble runs for 1 ns are used to further equilibrate the structures at 300 K. Then, the amorphous part is insert to the unpoled crystalline part and the whole structure is relaxed at an NPT ensemble for 1 ns to get an unpoled semi-crystalline structure. And an external electric field of 2 GV/m along y direction to the unpoled structure in MD simulation. With electric field applied, the structure is simulated in NPT ensembles at 300K and 1 atm for 100 ps, followed by NVE ensembles for 100 ps, and the structure is poled. Then the electric field is removed, followed by the same steps that is used to calculate thermal conductivities of unpoled and poled semi-crystalline structure. The integrals of HCACF along *y* direction are shown in Figure S17(c). The calculated thermal conductivity along y direction ($\kappa_y$) of unpoled PVDF is $0.19 \pm 0.01$ Wm$^{-1}$K$^{-1}$, and the calculated $\kappa_y$ of the poled structure is $0.28 \pm 0.03$ Wm$^{-1}$K$^{-1}$. To qualitatively analyze the effect of electric field, the relaxed structures of crystalline part before and after electric poling are shown in Figure S17(d) and S17(e), respectively. There is a significant discrepancy in inter-chain morphology between two structures. The orientations of molecular chains in unpoled PVDF along *x* and *y* directions are inconsistent, resulting a disorder inter-chain structure. While in poled PVDF, all molecular chains orient consistently along the direction of electric field, generating a more ordered inter-chain structure. The improvement of inter-chain order would suppress phonon scattering, resulting a larger $\kappa_y$ in poled PVDF.

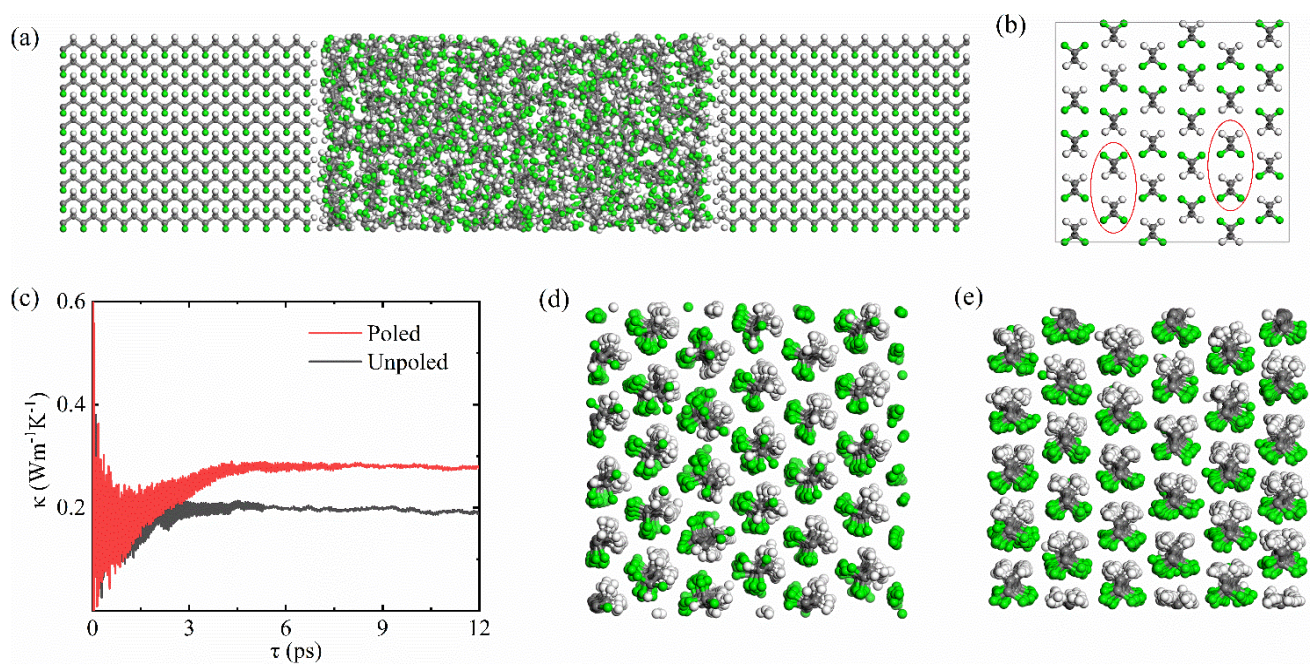

**Figure S17.** (a) Initial structure of semi-crystalline PVDF; (b) side view of the initial unpoled crystalline part; (c) Integrals of HCACF along y direction at 300 K. Relaxed structures of crystalline part (d) before and (e) after electric poling